\documentclass[12pt,preprint]{aastex}

\slugcomment{To appear in ApJ}

\shorttitle{Spectroscopy of low mass members  in Chamaeleon II}
\shortauthors{Barrado y Navascu\'es \& Jayawardhana}

\def\lsun{L_{\sun}}

\newcommand{\msun}{$M_{\odot}$}

\begin{document}

\title{A new Classical T Tauri object at the sub-stellar 
boundary in Chamaeleon II}

\author{David Barrado y Navascu\'es}
\affil{Laboratorio de Astrof\'{\i}sica Espacial y F\'{\i}sica Fundamental,
INTA, P.O. Box 50727, E-2808 Madrid, SPAIN}
\email{barrado@laeff.esa.es}

\author{Ray Jayawardhana}
\affil{Department of Astronomy \& Astrophysics, University of Toronto, Toronto, 
Ontario M5S 3H8, CANADA}
\email{rayjay@astro.utoronto.ca}

\begin{abstract}
We have obtained low- and medium-resolution optical spectra of 20 candidate
young low-mass stars and brown dwarfs in the nearby Chamaeleon II dark cloud, using 
the Magellan Baade telescope. We analyze these data in conjunction with near-infrared 
photometry from the 2-Micron All Sky Survey. We find that one target, [VCE2001] C41, exhibits 
broad H$\alpha$ emission as well as a variety of forbidden emission lines. These 
signatures are usually associated with accretion and outflow in young stars and brown 
dwarfs. Our spectra of C41 also reveal \ion{Li}{1}  in absorption and allow us to derive a 
spectral type of M5.5 for it. Therefore, we propose that C41 is a classical T Tauri 
object near the sub-stellar boundary. Thirteen other targets in our sample have 
continuum spectra without intrinsic absorption or emission features, and are difficult 
to characterize. They may be background giants or foreground field stars 
not associated with the cloud or 
embedded protostars, and need further investigation. The six remaining candidates, with 
moderate reddening, are likely to be older field dwarfs, given their spectral types, 
lack of lithium and H$\alpha$. 
\end{abstract}


\section{Introduction}

The southern star-forming region in Chamaeleon consists of three major dark clouds --designated 
Cha~I, Cha~II and Cha~III-- and 
a handful of smaller structures (Schwartz 1977; Boulanger et al. 1998). 
Given its relative proximity (160--180 pc; Whittet et al. 1997), youth ($\sim$1 Myr) and
intermediate galactic latitude ($b \sim -17 \deg$), 
this cloud complex is well suited for investigations of low-mass stellar and sub-stellar
populations. Cha~I has been the focus of a variety of studies (Luhman 2004 and 
references therein) whereas Cha~II and Cha~III remain relatively unexplored. 

The interest in Cha~II, in particular, is likely to grow in the near future because it will 
be surveyed during the course of a Spitzer Legacy program (Evans et al. 2003). Schwartz (1977)
and Hartigan (1993) identified a number of pre-main-sequence stars with H$\alpha$ emission
in this region. Near-infrared and InfraRed Astronomy Satellite (IRAS) data confirmed these 
sources likely harbor circumstellar material (Whittet et al. 1991; Prusti et al. 1992; Larson 
et al. 1998). A Class I protostar, 
IRAS 12553-7651, with a probable molecular outflow and near-infrared nebulosity, has been found  
close to the cloud center (Olmi et al. 1997; Persi et al. 2003). A recent Infrared Space 
Observatory (ISO) study revealed four new candidate young stellar objects, including a very 
low luminosity ($\ge$ 0.01 $\lsun$) source with a mid-infrared excess (Persi et al. 2003). 
Alcal\'a et al. (2000) detected 40 X-ray sources in Cha~II in a ROSAT pointed observation; only
14 of these sources coincided with previously known young stellar objects. These authors suggest
that Cha~II contains fewer weak-line T Tauri stars than classical T Tauri stars, compared to Cha~I, 
and thus may be in an earlier evolutionary stage than the latter. 

Using $I, J, K_s$ data from the DEep Near Infrared Survey of the southern sky (DENIS; Epchtein 
et al. 1997), Vuong, Cambr\'esy \& Epchtein (2001, hereafter VCE2001) identified 51 candidate low-mass stars
and brown dwarfs in Cha~II. Comparison with theoretical evolutionary tracks suggest that their 
photometrically-selected candidates have masses $<$ 0.2 \msun~ and ages 1--10 Myr. However, 
the true nature of these candidates can only be determined through spectroscopy. Here we present
optical spectra of 20 of the DENIS candidates, and use these data to derive their
spectral types and to evaluate their evolutionary status and cluster membership.

\section{Observations}
We obtained spectra of 20 Cha~II targets at low-resolution (2003 March 9) and four of 
them also at medium-resolution (2003 March 10 and 11) with the Boller \& Chivens (B\&C) 
optical spectrograph on the Magellan Baade 6.5-meter telescope. The spectral resolutions 
were R=620 and R=2600, as measured in the comparison arc lamps, using the 300 l/mm and 
1200 l/mm gratings. The spectral coverages were  5200--9600 \AA{ } and 6200--7800 \AA~ 
respectively. The data were reduced using standard techniques within IRAF\footnote{IRAF 
is distributed by National Optical Astronomy Observatories, which is operated by the 
Association of Universities for Research in Astronomy, Inc., under contract to the 
National Science Foundation, USA}.

In the case of the low resolution spectra, typical exposure times were of the order
 of 300 seconds. For the fainter objects, we collected several exposures of this exposure time 
and co-added them after the reduction process. For the medium resolution 
spectra, we collected 3--4 exposures of 900 seconds each.  

On the same observing run, we observed a large number of late K and M 
stars of III, IV and V luminosity classes, and several 
spectrophotometric standard stars. These spectra were used to calibrate 
the data and to derive spectral types. During this run, we also 
observed members of other young clusters and stellar associations, such as 
TW Hya association, IC2391 cluster and $\lambda$ Orionis cluster. Further 
details can be found in 
Mohanty, Jayawardhana \& Barrado y Navascu\'es (2003);
Barrado y Navascu\'es, Mohanty \& Jayawardhana  (2004);
Barrado y Navascu\'es et al.  (2004); and
Barrado y Navascu\'es, Stauffer \& Jayawardhana  (2004).

%%%%%%%%%%%%%%
%HIP66886----K7Mag2003N1_090   
%HIP65133----K8Mag2003N1_093   
%LTT-4364-------Mag2003N1_086   
%LTT-4816-------Mag2003N1_088   
%GJ223.1-----M0Mag2003N2_025   
%HIP66675----M0VMag2003N1_095   
%HIP66338----M1IIIMag2003N1_096   
%GJ173-------M1Mag2003N2_023   
%HIP65859----M1VMag2003N1_098   
%HIP72689----M2IIIMag2003N1_099   
%GJ2043A-----M2Mag2003N2_026   
%HIP72944----M2VeMag2003N1_101   
%HIP74005----M3IIIMag2003N1_103   
%HIP67155----M3VMag2003N1_104   
%HIP69981----M4-5IIIMag2003N1_106   
%HIP9981-----M4-M5IIIMag2003N2_072   
%HIP66077----M4Ve-sMag2003N1_107   
%HIP71868----M5IIIMag2003N1_110   
%GJ402-------M5?Mag2003N1_054   
%HIP025578---M5Mag2003N1_018   
%HIP70669----M6IIIe-M8eMag2003N1_113   
%GJ406-------M6Mag2003N2_037   
%HIP68137----M7Mag2003N2_071   
%vB08--------M7Mag2003N3_057   
%HIP63642----M8IIIMag2003N2_062   
%HIP763642---M8IIIMag2003N1_112   
%LHS2243-----M8Mag2003N1_057   
%LHS2397a----M8Mag2003N3_054   
%bri-0337-35-M9Mag2003N1_016   
%%%%%%%%%%%%%%%%%%

In addition, we have compiled near-infrared data --$JHKs$ photometry-- from 
the 2MASS All Sky Survey (Cutri et al. 2003). These values are listed, 
together with previous optical-infrared data --$IJK$-- from DENIS, in Table 1.

\section{Results and Discussion}
We use these data to re-estimate the reddening values, 
 derive spectral types for the sample objects and evaluate their nature
and possible membership in Cha~II. Figure 1 shows the distribution of these 
targets on the sky against a contour plot of 100 $\mu$m emission from IRAS, as well
as the distribution of X-ray sources and Cha~II members of IR classes I, II and III
(after Alcal\'a et al. 2000).

%Dots are Denis candidates listed in Vuong, Cambr\'esy, Epchtein (2001),
%solid triangles denote those candidates which have a continuum spectrum,
%those with small veiling appear as solid circles, non-members (dwarfs)
%are indicated with crosses and those having forbidden lines 
%are  indicated with large open circles.

\subsection{Color-Magnitude and Color-Color Diagrams: Reddening}
Figures 2, 3 and 4 display color-magnitude and color-color diagrams (CMD and 
CCD), respectively, for the sample analyzed in this paper. 
The first set of diagrams (Figure 2) suggests that the VCE2001 study contains 
a large number of young substellar objects. However, it is clear that 
there are significant differences in reddening (Figure 3 and 4). VCE2001, in their 
discovery paper, presented a reddening map and derived, based on it, the individual 
reddening for each object. These values are listed in Table 1. On the other hand, 
the IRAS contours plotted in Figure 1 clearly show that there is differential 
reddening, both  for those objects within the Cha~II cloud and for background 
objects located behind it. 

Figure 4 displays the detail for the ($J-H$) versus ($H-Ks$) diagram --based on 
2MASS data. We show a comparison between other Cha~II members of different IR class
and the VCE2001 sample in panel (a), 
whereas panel (b) contains the sample of X-ray emitters (data from
Alcal\'a et al. 2000). Note that most Class II objects 
(in accreting phase, Lada \& Wilking 1984; Shu et al. 1987)
are located over the CTT locus. Only one Class I object appears in the 2MASS database, and 
it is located in the same position as VCE2001 candidate members with 
large reddening values. We have been able to identify only few X-ray sources
 from Alcal\'a et al. (2000) in 2MASS, and they are located moderately close to the main sequence. The two diagrams suggest that these populations are different
(i.e., the VCE2001 sample may contain a large number of spurious members).

We show the subsample of the objects with continuum spectra (see next section) in  Figure 4c and 4d.
 This subsample is characterized by large reddening towards each of the objects.
 In panel 4c, we display the location 
of this subsample and the reddening vector based on the  VCE2001 data (i.e, the 
location of the arrow tip indicates the color of each object after reportedly been 
corrected for interstellar reddening). As visual inspection indicates, the dereddened 
objects  follow neither the main sequence nor the classical T Tauri locus. We have 
procceded to derive again the reddening vector by forcing each object to lie
on the main sequence (or on the point closest to it). Figure 4d shows the subsample
with the new reddening vectors. These values are also included in Table 1. 
The comparison of the interstellar absorption derived by VCE2001 and by us
indicates that there are important differences.

\subsection{Spectral types and membership}

\subsubsection{Subsample with large reddening}

Figure 5a shows the spectra of the subsample with large reddening. All of them are 
continuum spectra, and only contain telluric features. The spectral slope of these 
objects, after correction of the intrumental response but prior to dereddening, is 
similar to the slope of mid-M spectral type. However, if we apply the 
extinction derived by VCE2001, the slope is closer to early M for most of them.
Finally, if we apply our own, newly derived reddening, they are closer to K7-M0.

These  13 featureless objects are difficult to characterize. Featureless spectra in 
the optical are characteristic of embedded objects or classical T Tauri stars with 
very strong veiling, in which the contribution from circumstellar dust is larger 
than the photospheric emission from the central object (Lada \& Wilking  1984; 
Adams et al. 1987; Bertout et al. 1988; Bertout  1989; Appenzeller \& Mundt 1989). 
From  Figure 1, it appears that these 13 sources are associated with the denser parts 
of the Chamaeleon II cloud. This might suggest that they are embedded stellar and sub-stellar 
objects, still in their Class I phase. At least one confirmed Class I protostar,  
IRAS 12553-7651, with a probable molecular outflow and near-infrared nebulosity had 
been previously identified close to the center of Cha~II (Olmi et al. 1997; Persi 
et al. 2003). However, other optically visible Class I objects, such as Sz102 in 
Lupus (Comer\'on et al. 2003) and some  Taurus members (Kenyon et al. 1998),
show intense  H$\alpha$ emission and forbidden lines. This is not the case for the  
13 continuum sources in our sample. Therefore, until further evidence is collected 
--near-infrared spectroscopy or mid-infrared photometry-- it may be safer to assume 
that these objects are probable background, red giants, not associated with the
Cha~II cloud. 
Note that Luhman (2004), in his study of Cha~I, has found that up to 50\% of the 
candidate members can be foreground field stars and background giants. 
In addition, Figure 1 shows that the distribution of Cha~II members (Class I, II and III
objects, together with X-ray sources from Alcal\'a et al. 2000) is different from 
that for VCE2001 candidates. In the first case, they are aligned with the bar of dense material, with a SE-NW orientation, whereas the other two groups, composed of probable members of the star forming region, are concentrated North-East of this area.

\subsubsection{Subsample with small reddening}

For the other seven objects in our sample, which have moderate reddening, 
we derived spectral types by direct comparison with a large sample of template 
stars. Estimated uncertainty is about half a sub-class. Figure 6 shows the 
low-resolution spectra. Panel (a) consists of those seven candidate members, 
after instrumental response correction but without taking the reddening into account.
Panel (b) displays the spectra after we have applied the reddening corrections, whose 
values were obtained in a manner similar to that described in section  3.1 and 
illustrated in Figure 4b. Our derived spectral types are listed in Table 1. 

Figure 7 shows medium-resolution spectra of four targets with moderate reddening. 
Again, we have derived their spectral type and luminosity class in comparison to 
template spectra. Comparison of the gravity sensitive \ion{K}{1} doublet at 7700 \AA{ } 
is shown in Figure 8. Three --namely  [VCE2001] C01, C03, and C15--
 out of the four exhibit strong 
potassium absorption and are likely foreground dwarfs. The non-detection of \ion{Li}{1}6708
 \AA{ } in [VCE2001] C01 and C03  also suggests that these are probably older field stars 
(Figure 9). Lithium depletion is not so clear in the case of [VCE2001] C15  
due to poorer  S/N in its spectrum. The fourth target observed at medium-resolution, 
[VCE2001] C41 (hereafter C41), 
clearly exhibits Li absorption as well as a variety of prominent emission lines 
including H$\alpha$, \ion{He}{1} 6678 \AA, \ion{[O]}{1}, \ion{[N]}{2}, and \ion{[S]}{2}, 
and is likely a  classical T Tauri cloud member (see section 3.3). 

Three other candidates with low reddening, for which we only have low-resolution 
spectra --[VCE2001] C02, C06, and C07--
 do not show H$\alpha$ in emission. Therefore, 
they seem to be old, field M dwarfs and can be classified as probable non-members. 
These three, as well as two of the three other likely non-members with medium-resolution
spectra, appear to be located well outside the peaks in the IRAS 100$\mu$m map 
(Figure 1).  

In summary, of the 20 targets we observed, six are likely field M dwarfs (though we 
are less certain that [VCE2001]  C15 is a non-member since lithium depletion is not clearly seen), 
one is a classical T Tauri star and the other 13 with continuum spectra are probably 
heavily reddened  background giants, seen through the cloud.

\subsection{[VCE2001] C41: accretion and outflow near the sub-stellar boundary}
Figure 10 shows the medium- and low-resolution spectra of [VCE2001] C41 while Table 2 lists the 
equivalent widths of various lines seen at medium-resolution. The detection of \ion{Li}{1} 
in absorption implies a young object. Assuming it is a member of the Cha~II group, the
M5.5 spectral type we derive places [VCE2001] C41 at or near the sub-stellar boundary at an age
of a few million years. The H$\alpha$ emission line is broad, with a full width at 
half-maximum and at 10\% of maximum intensity--as measured in the medium-resolution spectrum-- 
of $\sim$285 and $\sim$500 km/s, respectively  
(in any case, much broader than the spectral resolution of $\sim$120 km/s, FWHM). 
We also detect a prominent emission line of \ion{He}{1} (6678 
\AA) and the \ion{Ca}{2} infrared triplet. 
These signatures are characteristic of classical 
T Tauri stars and young brown dwarfs undergoing accretion (e.g., Jayawardhana, Mohanty \& 
Basri 2002, 2003; Barrado y Navascu\'es \& Mart\'{\i}n 2003).
 Furthermore, [VCE2001] C41 exhibits
significant excess in the near-infrared (Figures 7 and 8), consistent with a dusty
circum-(sub)stellar disk (e.g., Jayawardhana et al. 2003; Barrado y Navascu\'es et al. 
2003). 

The forbidden lines seen in [VCE2001] C41 --\ion{[O]}{1}, \ion{[O]}{2}, \ion{[N]}{2}, \ion{[S]}{2}--
 are commonly observed
in Harbig-Haro objects, Class I sources and some T Tauri stars and may arise in a jet
or outflow (Kenyon et al. 1998, Jayawardhana et al. 2002). Recently such forbidden 
line emission has been found in two other young objects (LS-RCrA 1 and Par-Lup3 4)
 near the sub-stellar boundary
in the R Coronae Australis  and Lupus 3 star forming regions
 (Fern\'andez and Comer\'on 2001; Comer\'on  et al. 2003;
 Barrado y Navascu\'es,  Mohanty \& Jayawardhana 2004). 
We note that [VCE2001]~C41 is 14.4 arcsec (about 2450 AU) North-West of IRAS 12554-7635,
 a Class II source, and the two may be related.

Taken together, the spectroscopic and photometric evidence makes a compelling case that
[VCE2001] C41 is a young object near the sub-stellar limit undergoing accretion as well as mass 
loss. This result is consistent with the recent finding that very low mass young objects,
including brown dwarfs, go through a T Tauri phase similar to their higher-mass 
counterparts (e.g., Jayawardhana, Mohanty \& Basri 2002, 2003; Jayawardhana et al. 2003; 
Barrado y Navascu\'es  et al. 2003; Barrado y Navascu\'es \& Mart\'{\i}n 2003;
Barrado y Navascu\'es,  Mohanty \& Jayawardhana 2004; Barrado y Navascu\'es 2004).
 So far, it appears that brown dwarf disks may
persist for timescales comparable to those of low-mass stars (Jayawardhana et al. 1999;
Mohanty, Jayawardhana \& Barrado y Navascu\'es 2003). Thus, the presence of an accreting
disk around [VCE2001] C41  only constraints its age to $\lesssim$ 10 Myr while the signatures of
a jet suggest it may be closer to $\sim$1 Myr. Based on the X-ray detection rates, 
Alcal\'a et al. (2000) suggest that weak-line T Tauri stars are less numerous than 
classical T Tauri stars in Cha~II, contrary to the situation in Cha~I and several 
other star-forming regions. This result, the authors point out, could be due to 
Cha~II harboring a younger population. 

\section{Concluding Remarks}
We have found a classical T Tauri object near the sub-stellar boundary, with 
signatures of disk accretion and outflow, associated with the Cha~II dark cloud. 
Thirteen targets in our sample have continuum spectra, and may be either 
unrelated background giants or embedded protostars. Six other targets
are likely field M dwarfs. Further infrared studies --ground-based 
near-infrared spectroscopy and Spitzer observations in the mid- and far-infrared-- 
are needed to investigate the nature of the embedded stellar and sub-stellar population
in Cha~II. More sensitive X-ray observations could reveal additional weak-line T Tauri 
stars missing from the current samples. 

\acknowledgements

We are grateful to the Magellan staff for outstanding support. We thank Beate Stelzer, Leonardo Testi, Lee Hartmann, Carlos Eiroa, John Stauffer and the anonymous referee for useful discussions and suggestions. This work was supported in part by NSF grant AST-0205130 to RJ. 
DByN is indebted to the Spanish ``Programa Ram\'on y Cajal'', 
PNAyA AYA2001-1124-C02 and PNAyA AYA2003-05355.

\setcounter{table}{0}
\begin{table*}
\footnotesize
\caption[ ]{{\bf a} The targets with large reddening.}
\footnotesize
\begin{tabular}{lcccllccccl}
\hline
[VCE2001] &  I         &    J     &    Ks      & Av     &&    J        &  H         &   Ks             & Av &   Sp.Type$^1$     \\
\cline{2-5}  \cline{7-10}                                                                                                     
    &  \multicolumn{4}{c}{Vuong et al. (2001)}    &&            \multicolumn{4}{c}{2MASS}             &                 \\
\hline
\hline                                                                                                                           
C14 & 17.81 0.17 & 14.90 0.14 & 13.19 0.18 &  2.1 &&  14.947 0.037 & 13.671 0.031 & 13.190 0.038 & 5.0  & C \\%M5.5 mid-M  
C20 & 16.86 0.11 & 14.26 0.11 & 12.65 0.14 &  2.7 &&  14.387 0.032 & 13.216 0.025 & 12.701 0.027 & 4.8  & C \\%M5.5 early-M
C21 & 17.99 0.18 & 14.90 0.14 & 13.28 0.19 &  2.8 &&  15.107 0.046 & 13.834 0.027 & 13.227 0.034 & 5.9  & C \\%M5.5 early-M
C22$\dagger$ & 16.37 0.09 & 13.87 0.10 & 12.40 0.11 &  2.5 &&  14.028 0.028 & 12.903 0.025 & 12.552 0.030 & 3.3  & C \\%M4.5 early-M
C26 & 16.80 0.10 & 14.23 0.11 & 12.74 0.14 &  2.6 &&  14.331 0.030 & 13.343 0.029 & 12.819 0.030 & 3.5  & C \\%M5.5 early-M
C28 & 16.04 0.07 & 13.44 0.08 & 11.74 0.09 &  2.6 &&  13.546 0.028 & 12.288 0.023 & 11.789 0.021 & 5.2  & C \\%M5.5 early-M
C30 & 15.45 0.06 & 12.27 0.07 & 10.14 0.07 &  2.7 &&  12.251 0.023 & 10.755 0.024 & 10.208 0.021 & 6.5  & C \\%M6.0 mid-M  
C31 & 15.18 0.05 & 12.11 0.07 & 10.13 0.07 &  2.7 &&  12.109 0.023 & 10.672 0.022 & 10.144 0.019 & 6.1  & C \\%M6.0 mid-M  
C38 & 15.96 0.07 & 13.07 0.08 & 11.13 0.08 &  2.3 &&  13.105 0.023 & 11.736 0.025 & 11.217 0.023 & 6.0  & C \\%M5.5 mid-M  
C52 & 17.34 0.14 & 13.98 0.10 & 11.71 0.10 &  2.8 &&  14.091 0.028 & 12.402 0.025 & 11.733 0.023 & 8.5  & C \\%M6.0 early-M
C54 & 17.10 0.11 & 14.75 0.12 & 13.03 0.17 &  1.8 &&  14.842 0.043 & 13.908 0.042 & 13.437 0.045 & 3.0  & C \\%M5.0 early-M
C69 & 17.62 0.18 & 14.75 0.12 & 12.81 0.15 &  0.8 &&  15.022 0.053 & 13.386 0.033 & 12.862 0.036 & 7.0  & C \\%M6.0 mid-M  
C70 & 17.81 0.20 & 15.04 0.14 & 13.29 0.19 &  0.7 &&  15.129 0.056 & 14.056 0.039 & 13.493 0.050 & 4.3  & C \\%M5.5 mid-M  
\hline
\end{tabular}
$\,$
$^1$ ``C'' stands for continuum or featureless spectrum. \\
$\dagger$ IRAS 12510-7631, SIMBAD database.
\end{table*}

\setcounter{table}{0}
\begin{table*}
\footnotesize
\caption[ ]{{\bf b} The targets with low reddening.}
\begin{tabular}{lcccllccccl}
\hline
[VCE2001]&  I         &    J     &    Ks      & Av     &&    J        &  H         &   Ks             & Av &   Sp.Type       \\
\cline{2-5}  \cline{7-10}                                                                                                     
    &  \multicolumn{4}{c}{Vuong et al. (2001)}    &&            \multicolumn{4}{c}{2MASS}             &                 \\
\hline
\hline                                                                                                                           
C01 & 16.97 0.12 & 14.67 0.12 & 13.41 0.19 &  0.7 &&  14.777 0.040 & 14.019 0.047 & 13.759 0.059 & 0.8  &  M5.0V$^1$\\         
C02 & 17.77 0.44 & 14.75 0.13 & 13.43 0.17 &  1.1 &&  14.638 0.038 & 13.891 0.042 & 13.587 0.044 & 1.0  &  M4.5     \\         
C03 & 16.49 0.16 & 14.34 0.11 & 13.31 0.17 &  0.7 &&  14.405 0.035 & 13.694 0.032 & 13.426 0.035 & 0.6  &  M4.5V$^1$\\         
C06 & 17.32 0.12 & 14.70 0.12 & 12.98 0.14 &  1.0 &&  14.652 0.038 & 13.611 0.028 & 13.174 0.035 & 3.5  &  M5.5$^2$ \\         
C07 & 17.26 0.03 & 15.14 0.04 & 13.42 0.08 &  0.2 &&  15.264 0.053 & 14.301 0.049 & 13.989 0.062 & 2.5  &  M4.5$^3$ \\         
C15 & 16.88 0.10 & 14.41 0.12 & 12.96 0.16 &  1.9 &&  14.403 0.035 & 13.476 0.031 & 13.087 0.039 & 2.5  &  M5.5V$^{1,4}$\\  
C41$\dagger$ & 16.82 0.10 & 14.04 0.10 & 11.19 0.09 &  0.1 &&  14.446 0.051 & 12.566 --$^5$& 11.364 --$^5$&--$^6$&  M5.5$^1$ \\         
\hline
\end{tabular}
$\,$
$^1$ Observed also with the 1200 l/mm grating.\\
$^2$ M4.5 and M3.5  after derredening the spectra using the VCE2001 and our new value of the reddening. \\
$^3$ M4 and M2  after derredening the spectra using the VCE2001 and our new value of the reddening. \\
$^4$ M4.4 and M3.5  after derredening the spectra using the VCE2001 and our new value of the reddening. \\
$^5$ No error in the 2MASS database.\\
$^6$ Veiling.\\
$\dagger$ Associated to IRAS 12554-7635?
\end{table*}

\begin{table}
\caption{Equivalent widths of several lines in [VCE2001] C41. Note that the resolution is very different
for each night, since the data were taken with the 300 and 1200 lines/mm gratings, respectively. }
\begin{tabular}{lrr}
\hline
  Line             & March 10 & March 11 \\
                   & (\AA) &   (\AA) \\
\hline                            
\hline                            
  \ion{[O]}{1} 5577.35   & 17.3     &  --     \\
  \ion{[O]}{1} 6300.331  & 32.4     & 30.8    \\
  \ion{[O]}{1} 6363.79   &  7.7     &  7.3    \\
  \ion{[N]}{2} 6548      &  --      &  3.7    \\
  H$\alpha$ 6562.8       & 84.0$^1$ &108.5    \\
  \ion{[N]}{2} 6581      &  6.6     & 10.5    \\ 
  \ion{He}{1} 6678.15    &  --      &  1.0$^2$\\ 
  \ion{Li}{1} 6707.8     &  --      &  0.4    \\
  \ion{[S]}{2} 6717.0    &  6.3     &  5.9    \\ 
  \ion{[S]}{2} 6731.3    &  9.1     & 11.6    \\ 
  \ion{[O]}{2} 7319      &  4.7$^3$ &  4.5    \\ 
  \ion{[O]}{2} 7329      &  4.7$^3$ &  3.8    \\ 
  \ion{Ca}{2} 8498.0     &  2.3     &  --     \\
  \ion{Ca}{2} 8542.1     &  3.5     &  --     \\
  \ion{Ca}{2} 8662.1     &  2.5     &  --     \\
\hline
\end{tabular}
$\,$\\
$^1$ H$\alpha$ blended with \ion{[N]}{2} 6548.\\
$^2$ Lithium in  absorption.\\
$^3$ \ion{[O]}{2} 7319 \& 7329 blended together.
\end{table}

\setcounter{figure}{0}
%-----------------------------------------------------------
    \begin{figure*}
    \centering
    \includegraphics[width=7.8cm]{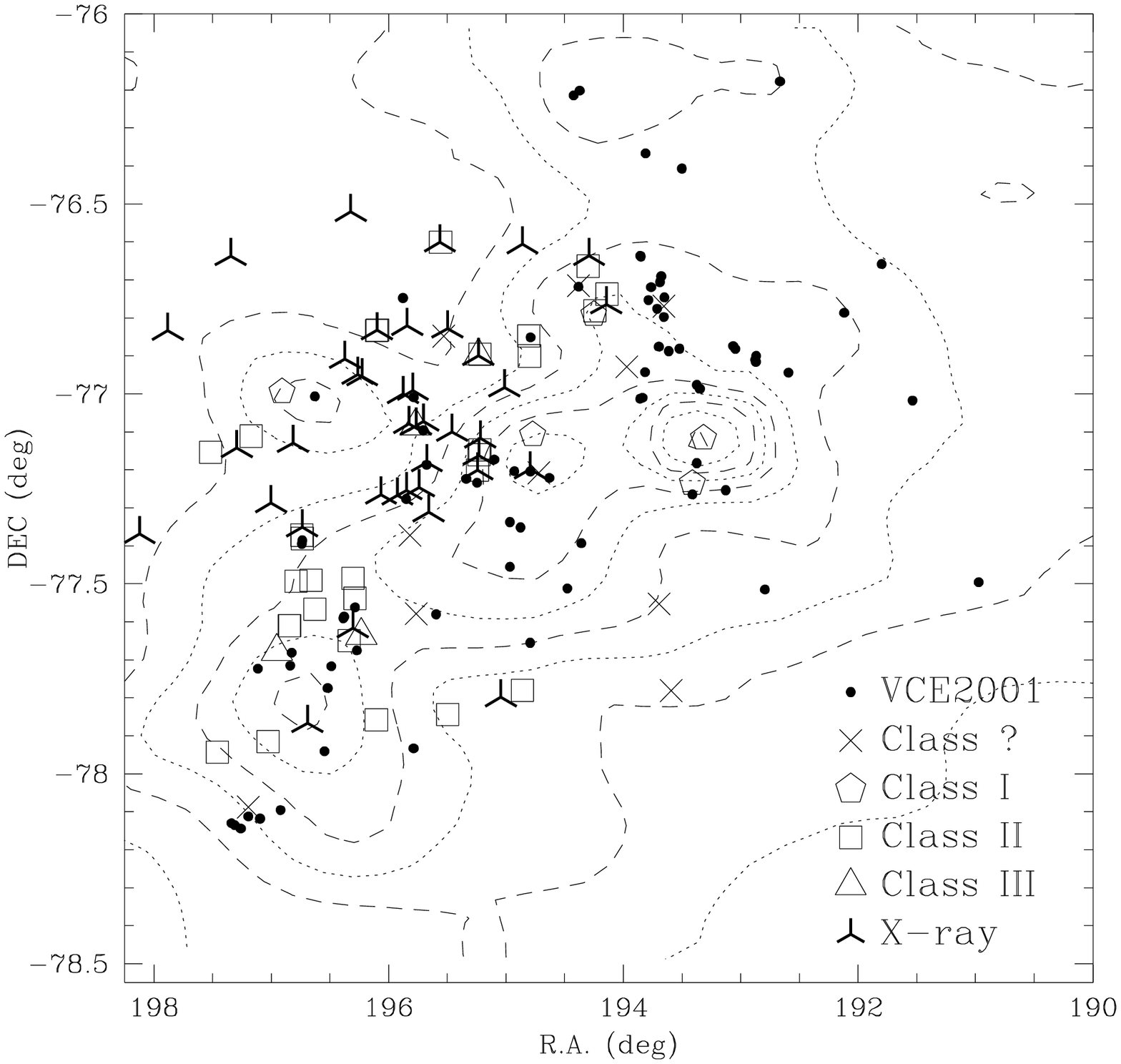}
    \includegraphics[width=7.8cm]{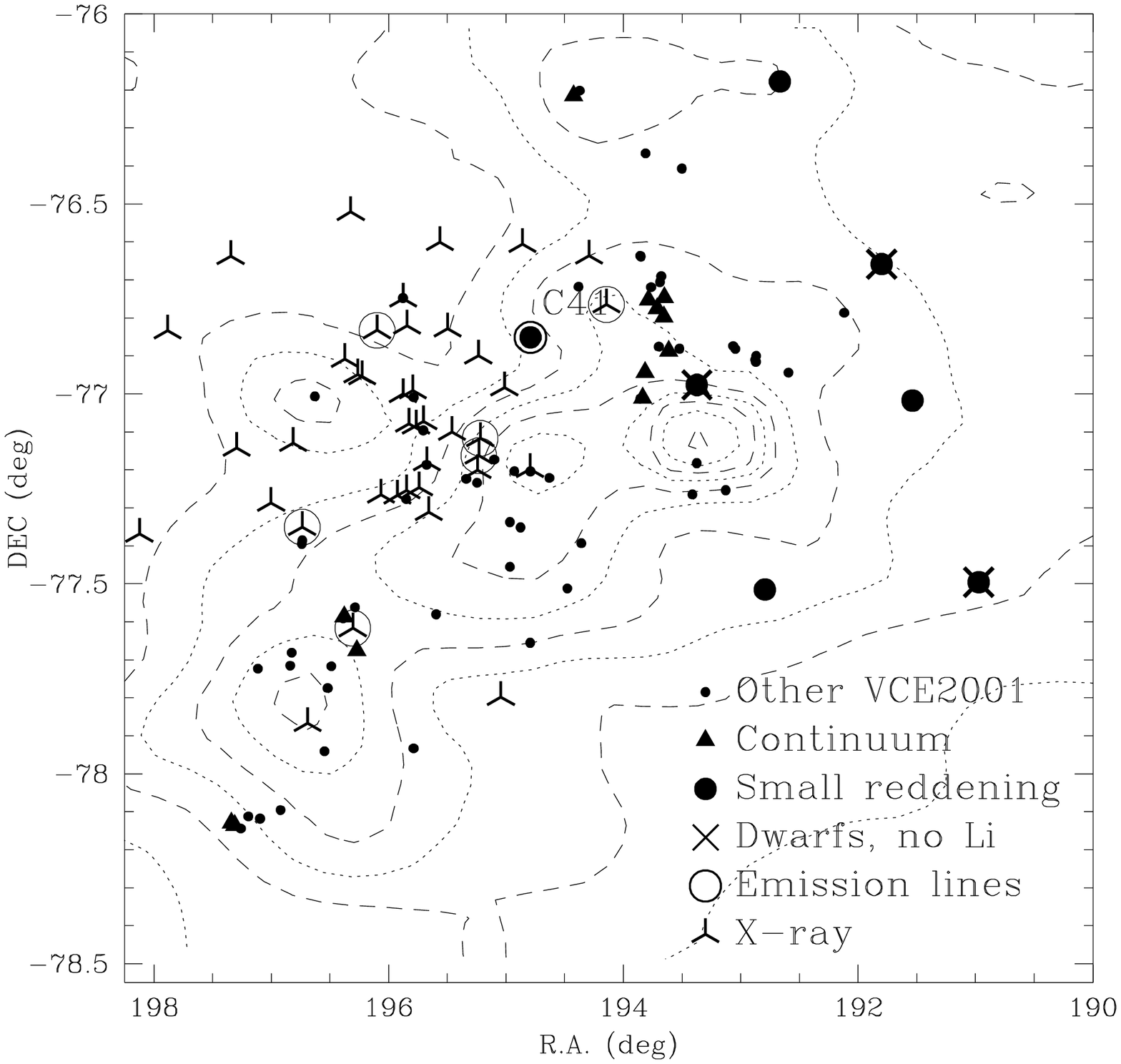}
 \caption{
Distribution of our targets on the sky.
The contours correspond to an IRAS map at 100 microns.
 Compare with figure 1 from Vuong et al. (2001) for the reddening map.
}
 \end{figure*}
%______________________________________________________________

\setcounter{figure}{1}
%-----------------------------------------------------------
    \begin{figure*}
    \centering
    \includegraphics[width=7.8cm]{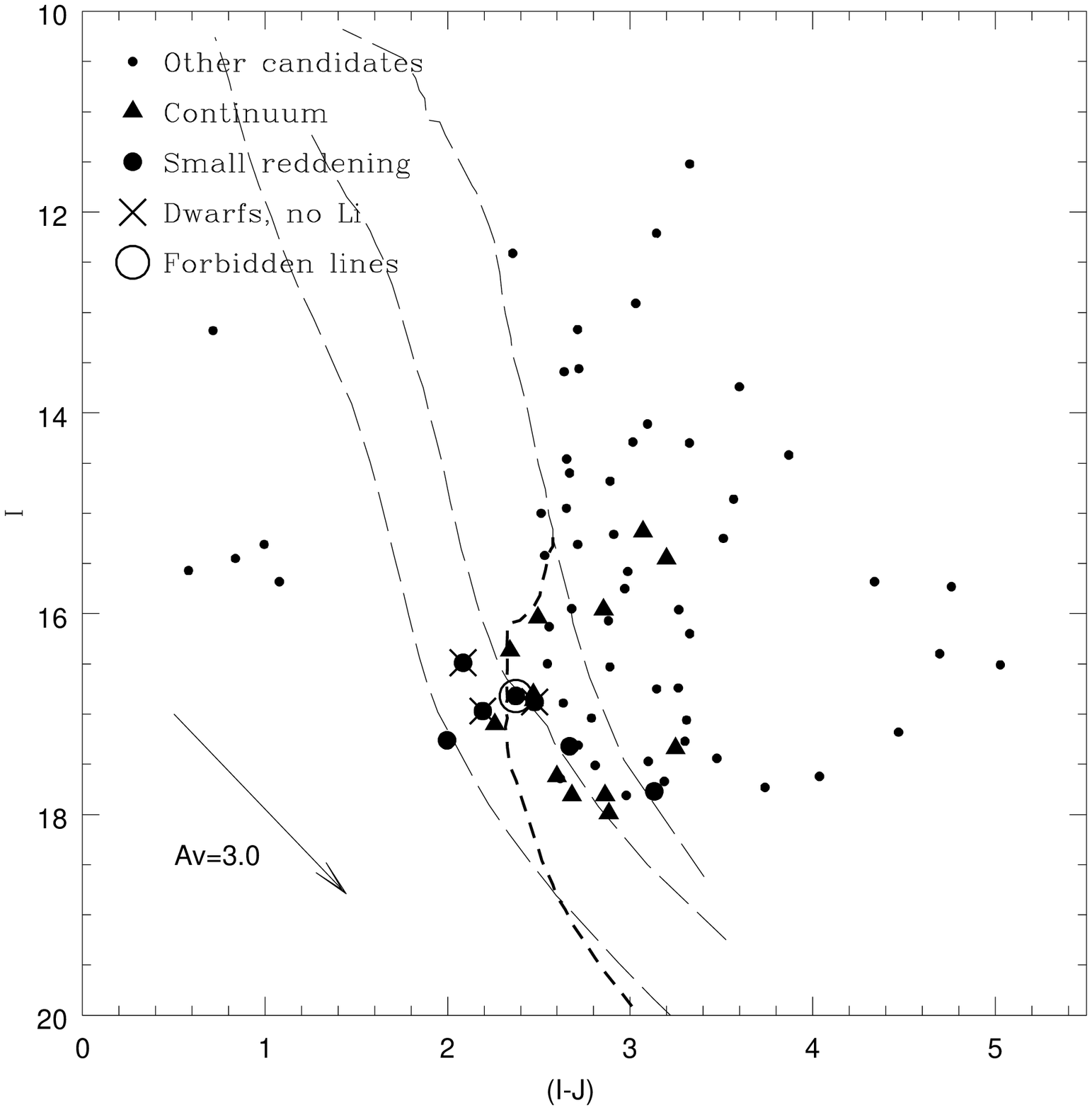}
    \includegraphics[width=7.8cm]{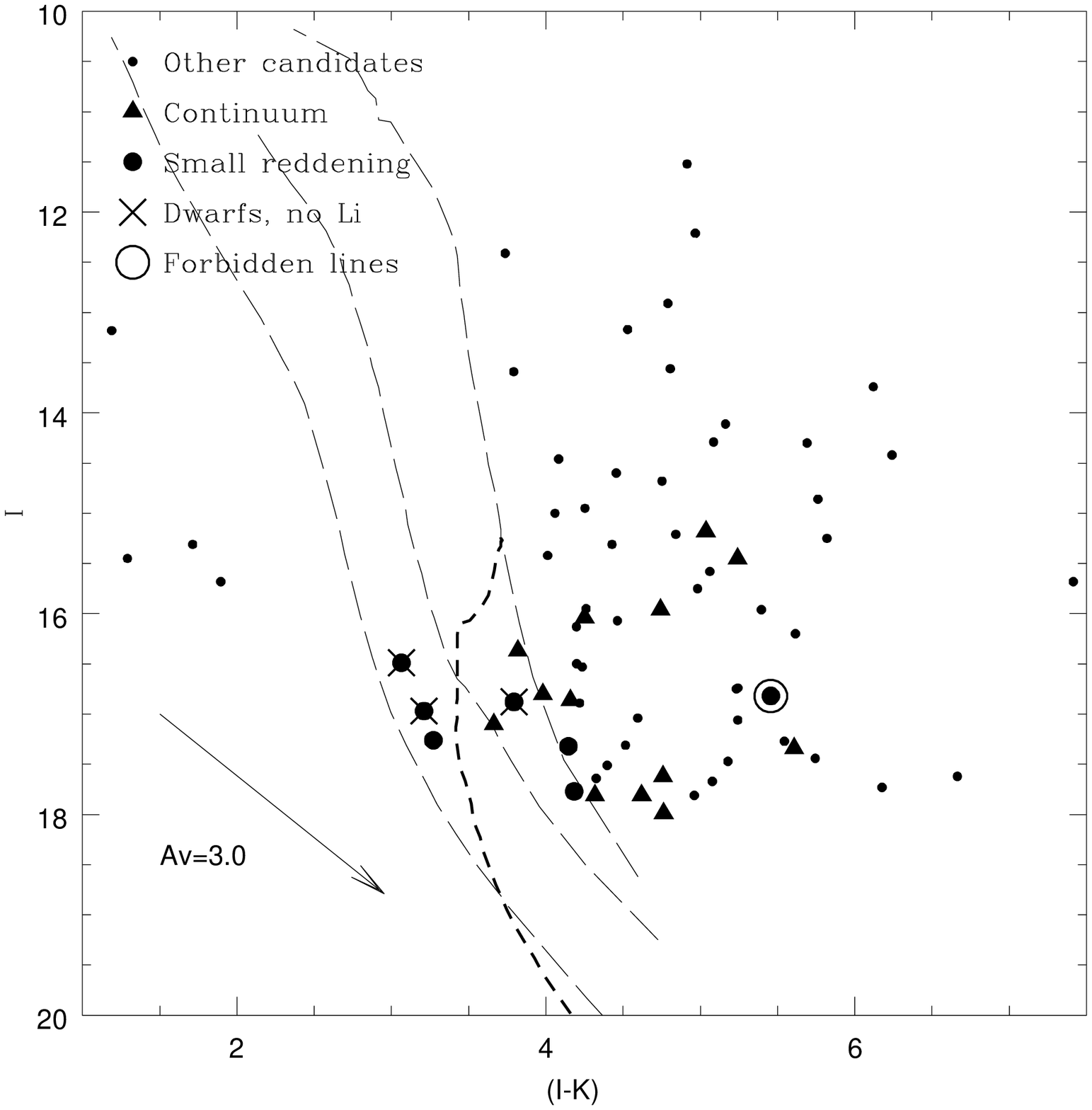}
    \includegraphics[width=7.8cm]{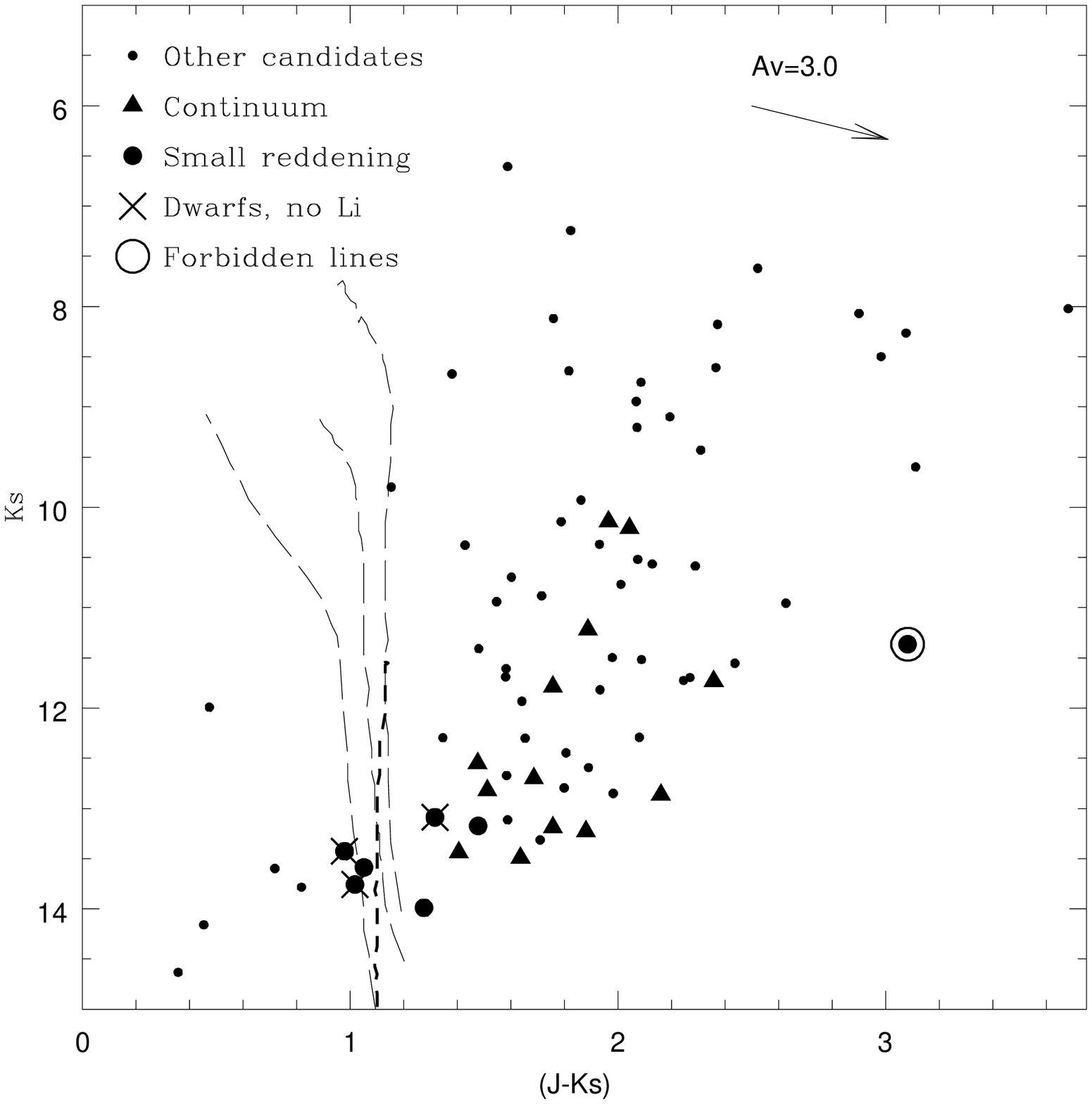}
 \caption{
Optical-infrared color-magnitude diagrams for
 low-mass stars and brown dwarfs  of  Cha~II cluster. 
Data from Vuong et al. (2001) and  2MASS All Sky Survey.
 Three isochrones  (1, 10 and 100 Myr) from  Baraffe et al$.$
 (1998) are included in the figure as long dashed lines.
The thick short dashed line correspond to a evolutionary track for 
0.072 M$_\odot$.  
}
 \end{figure*}
%______________________________________________________________      

\setcounter{figure}{2}
%-----------------------------------------------------------
    \begin{figure*}
    \centering
    \includegraphics[width=7.8cm]{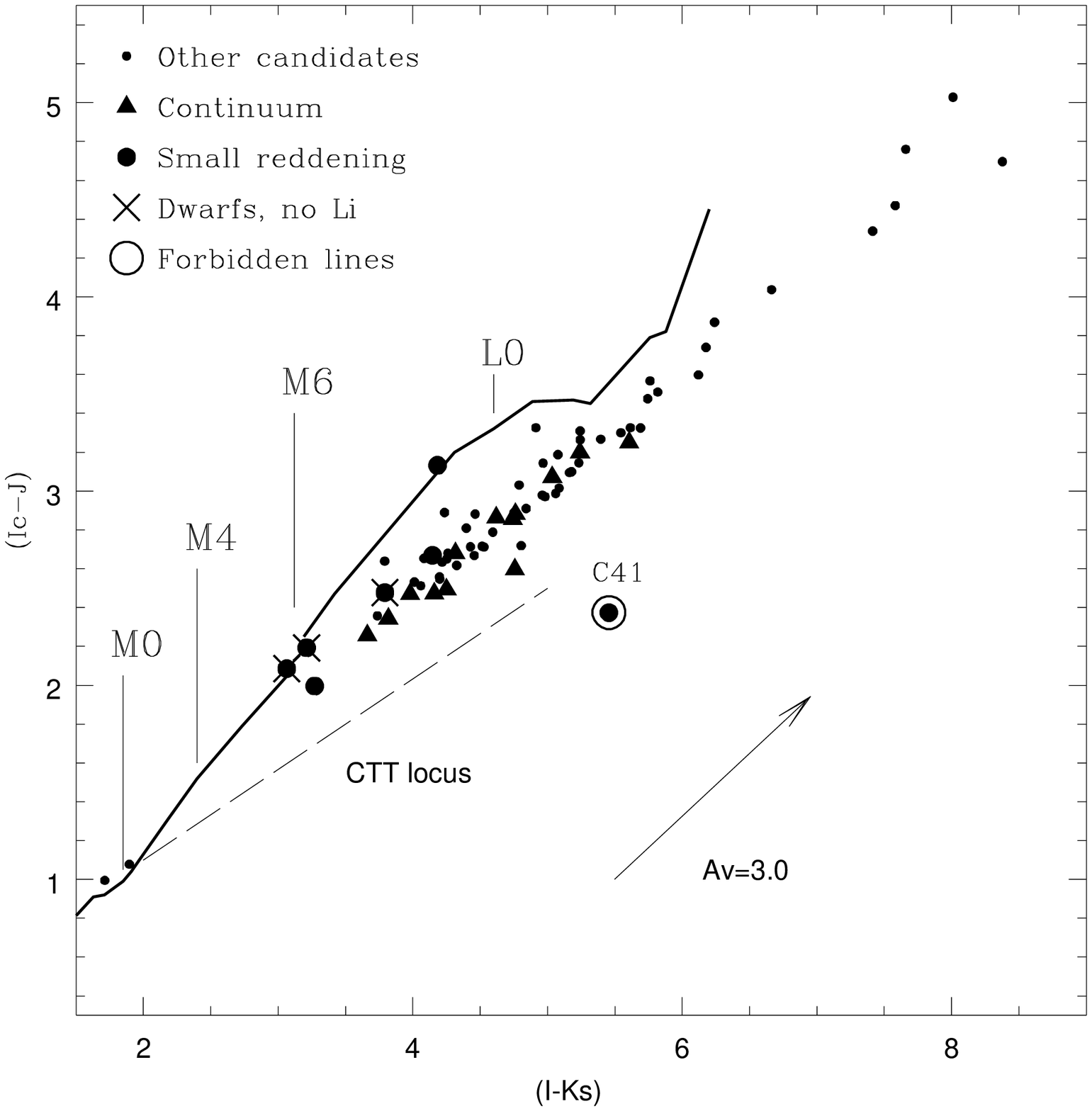}
    \includegraphics[width=7.8cm]{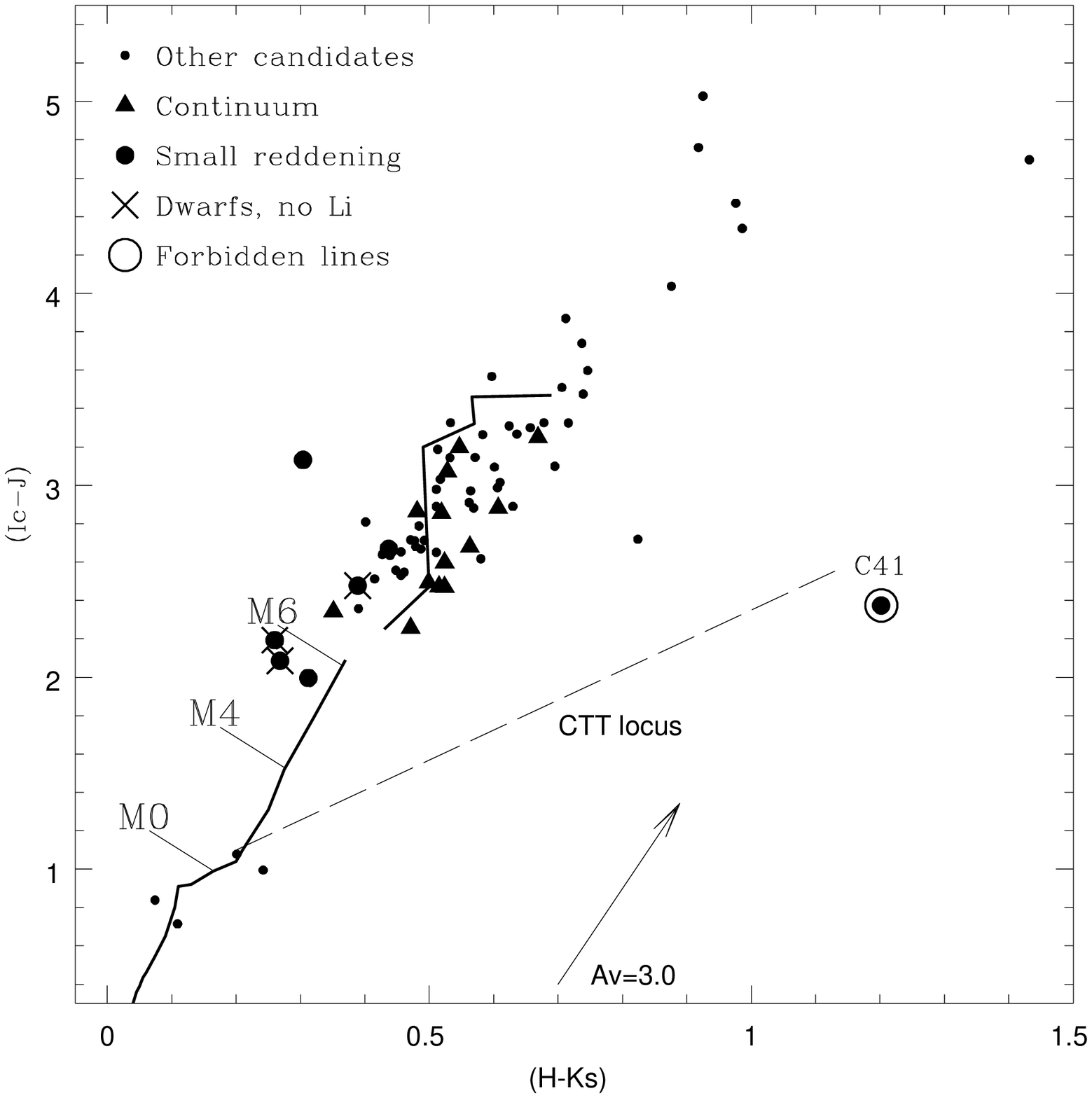}
 \caption{
Optical-Infrared color-color diagrams for Cha~II candadidates.
% Plus symbols indicate the position of classical 
%TTauri stars belonging to Orion
%stellar population (Herbig \& Bell 1988).
The thick-solid and dashed lines correspond to the locii of the main
 sequence stars (from Bessell \& Brett 1988; Kirkpatrick et al$.$
 2000; Leggett et al. 2001) 
and CTT stars (Meyer et al$.$ 1997 and this paper), respectively.  }
 \end{figure*}
%______________________________________________________________

\setcounter{figure}{3}
%-----------------------------------------------------------
    \begin{figure*}
    \centering
    \includegraphics[width=7.8cm]{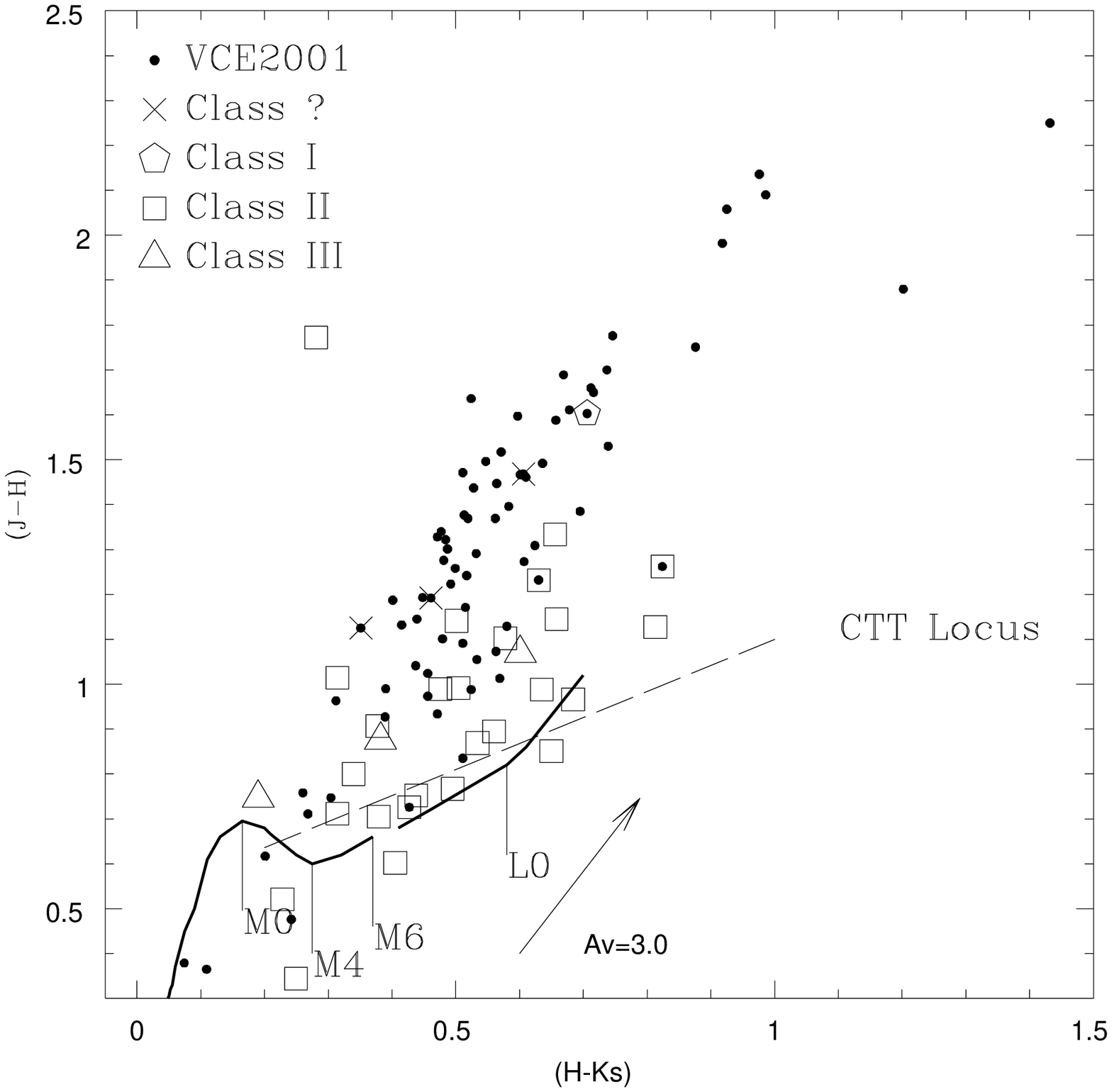}
    \includegraphics[width=7.8cm]{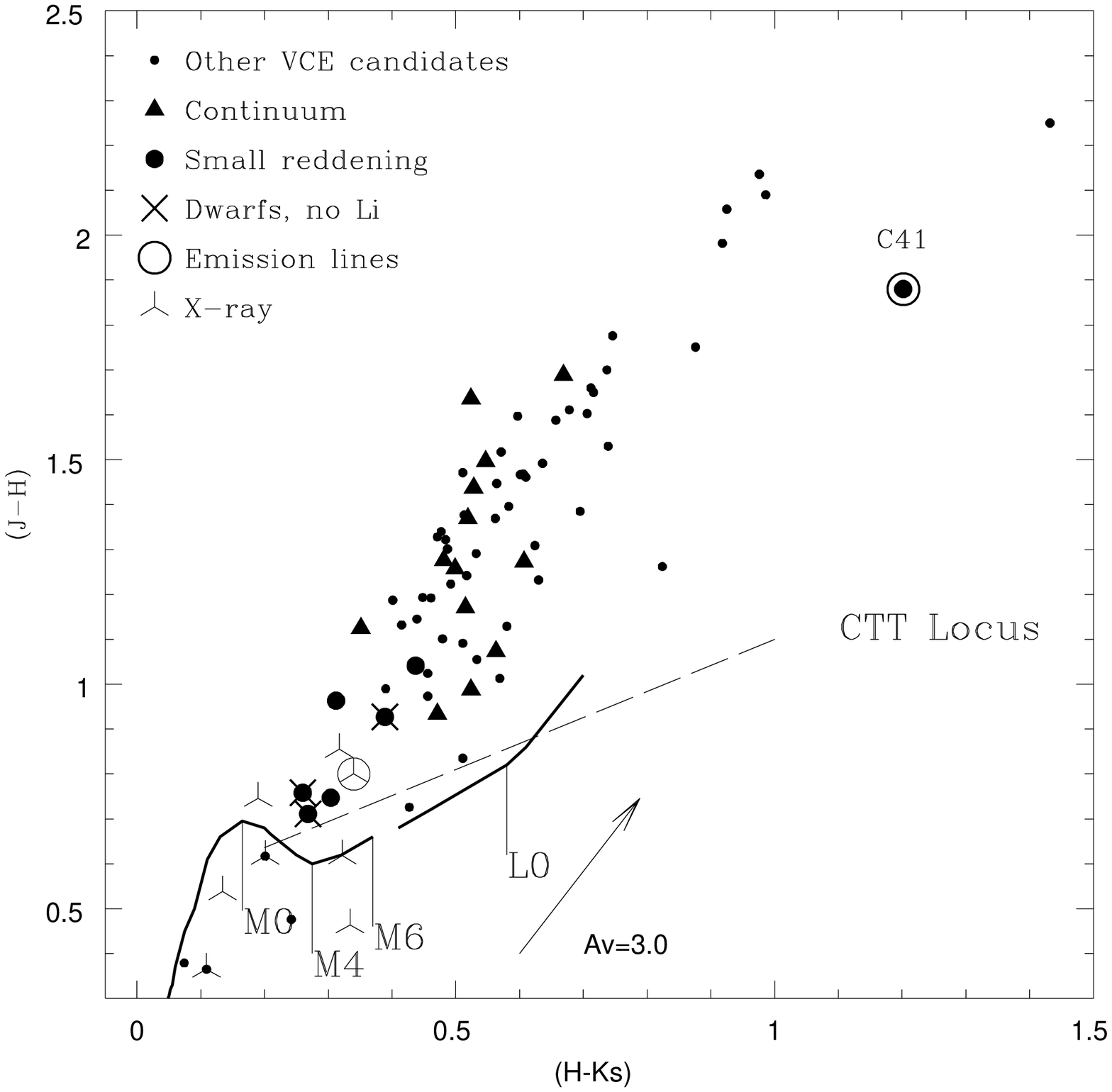}
    \includegraphics[width=7.8cm]{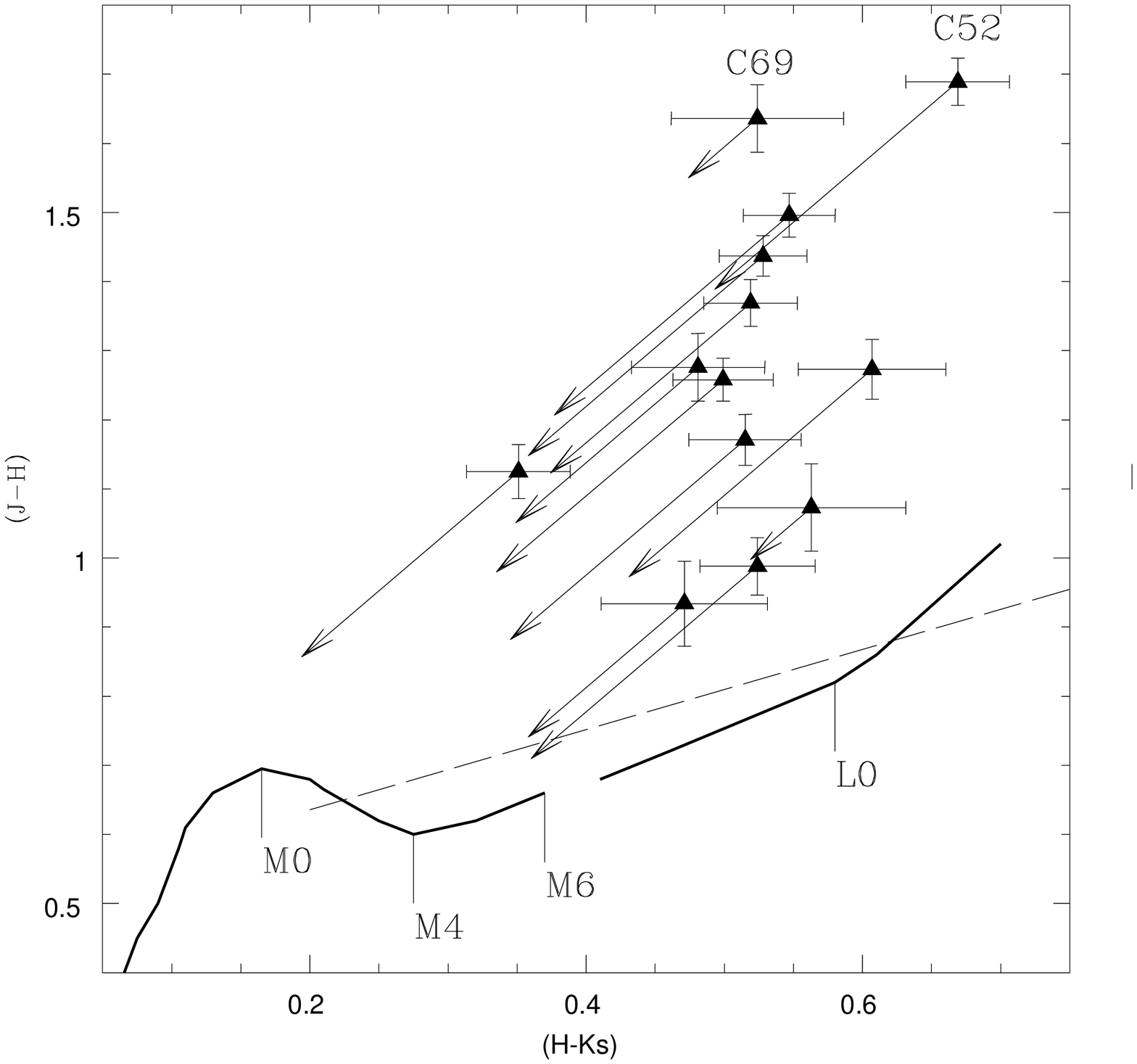}
    \includegraphics[width=7.8cm]{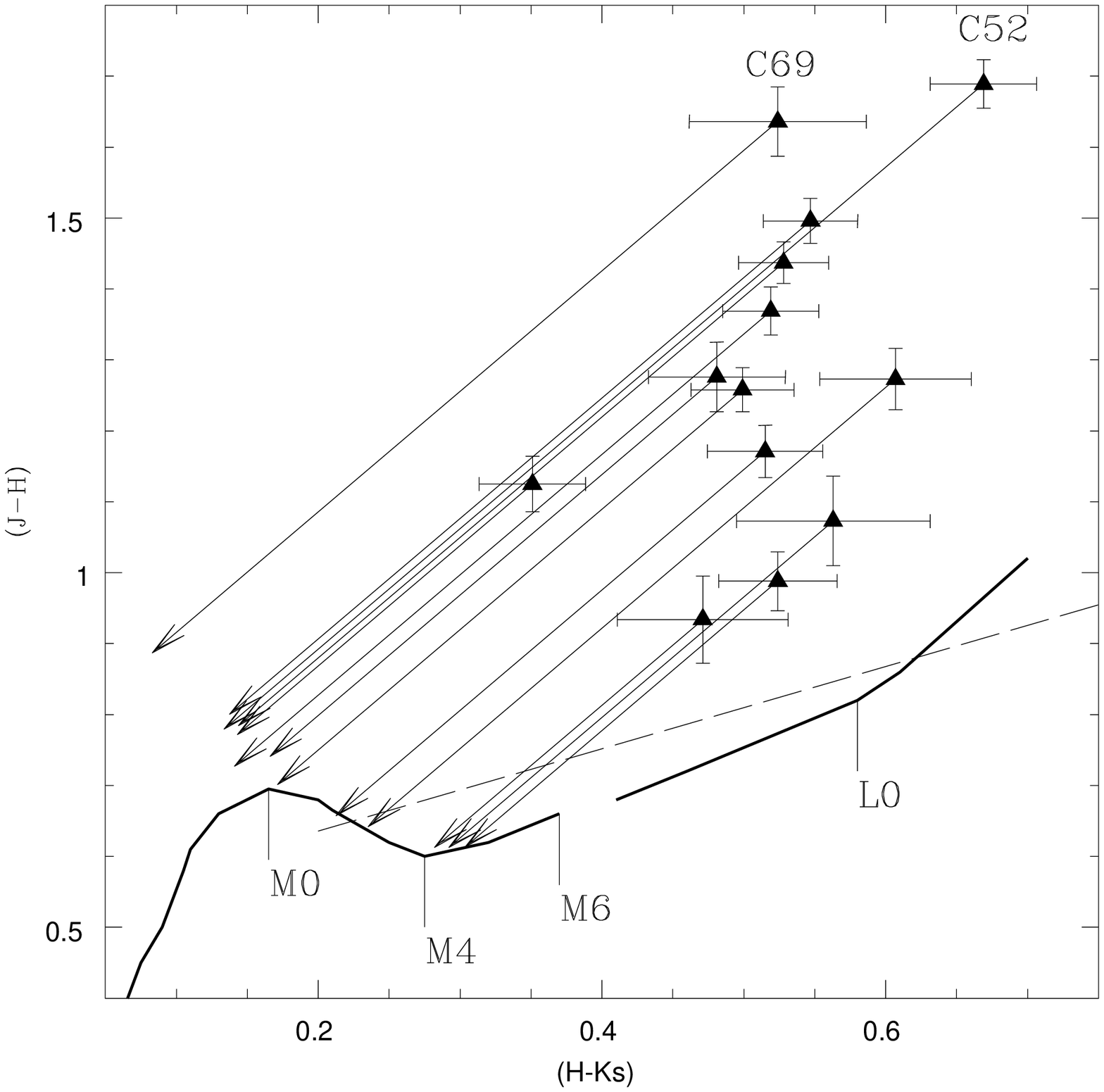}
 \caption{The ($J-H$) versus ($H-Ks$) CCD, displaying a
 zoom for the objects with continuum spectra.
Palen (a) and (b) display comparison with X-ray emitters and Cha~II members
of different classes (Alcal\'a et al. 2000), whereas
the effect of the reddening is illustrated in panels 
(c) and (d) --values from Vuong et al. (2001) and our own dereddening,
 based on 2MASS photometry and the Main Sequence locus.}
 \end{figure*}
%______________________________________________________________

\setcounter{figure}{4}
%-----------------------------------------------------------
    \begin{figure*}
    \centering
    \includegraphics[width=10.2cm]{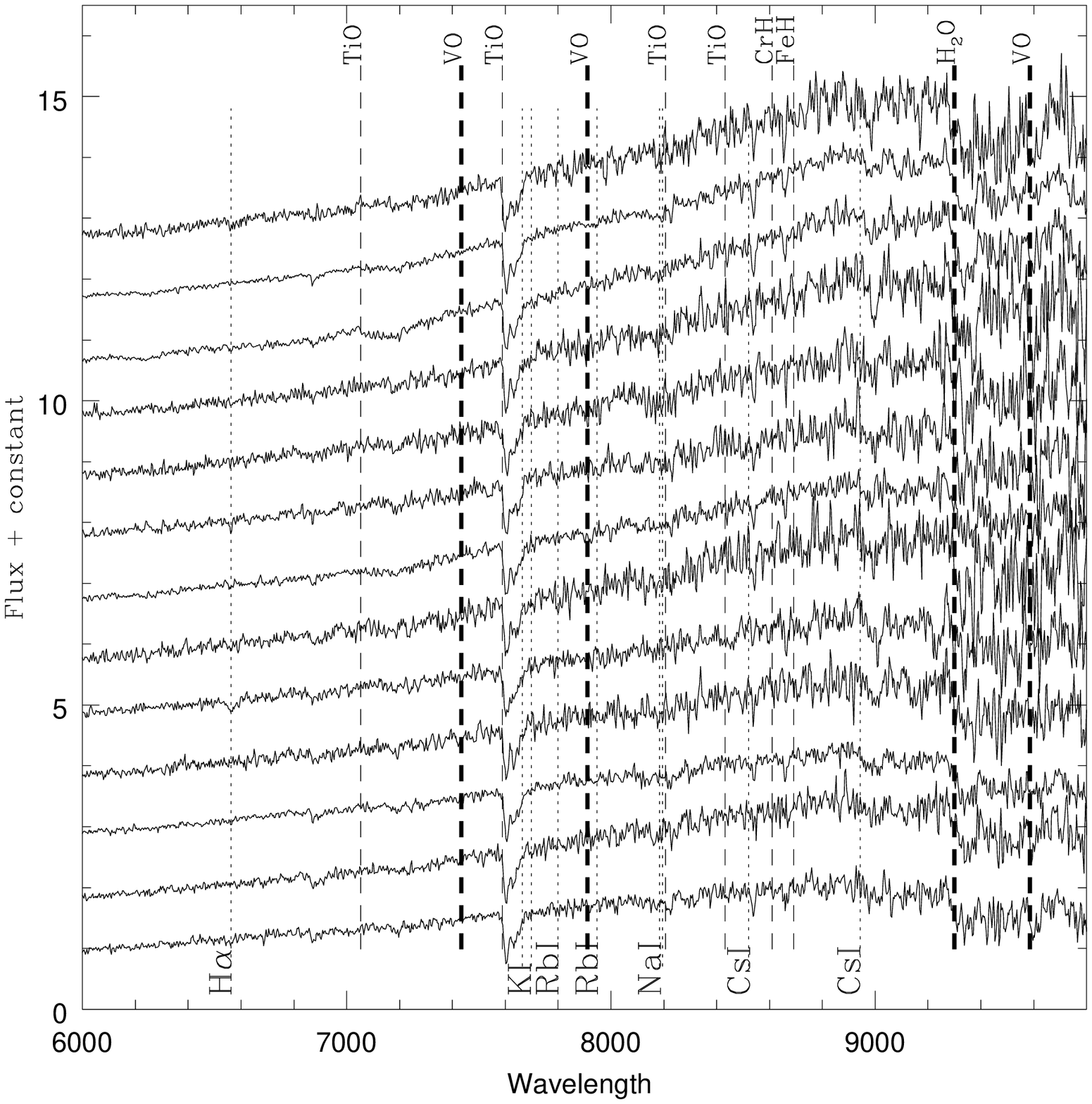}
    \includegraphics[width=10.2cm]{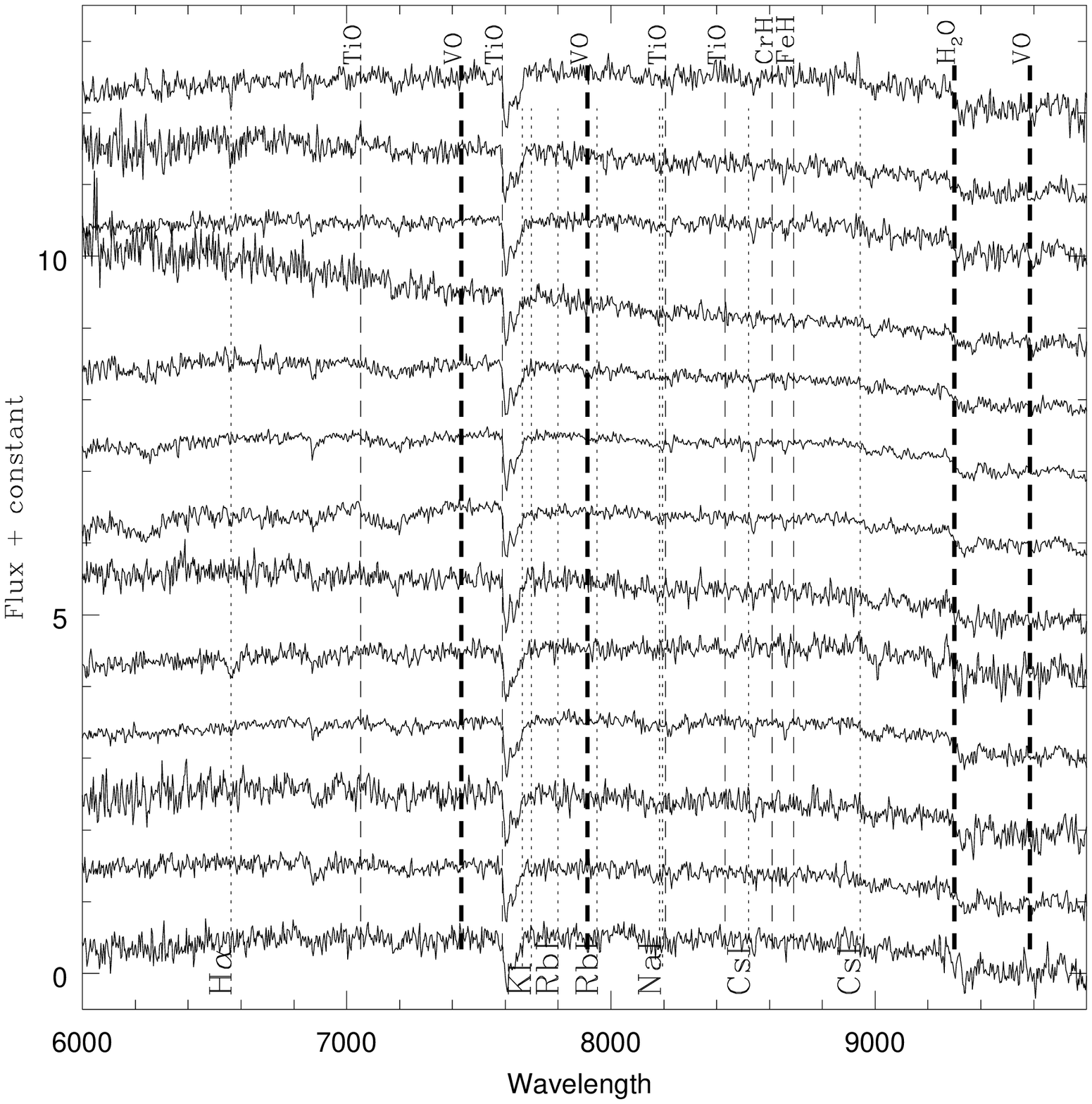}
 \caption{
Spectra taken with the 300 l/mm grating, corresponding to objects
 with very strong reddening. From bottom to top,
[VCE2001] C14, C20, C21, C22, C26, C28, C30, C31, C38, C52 --strong slope--, C54, C69 and C70.
{\bf a} Spectra prior dereddening.                    
{\bf b} Spectra after dereddening with our new values.
}
 \end{figure*}
%______________________________________________________________

\setcounter{figure}{5}
%-----------------------------------------------------------
    \begin{figure*}
    \centering
    \includegraphics[width=10.2cm]{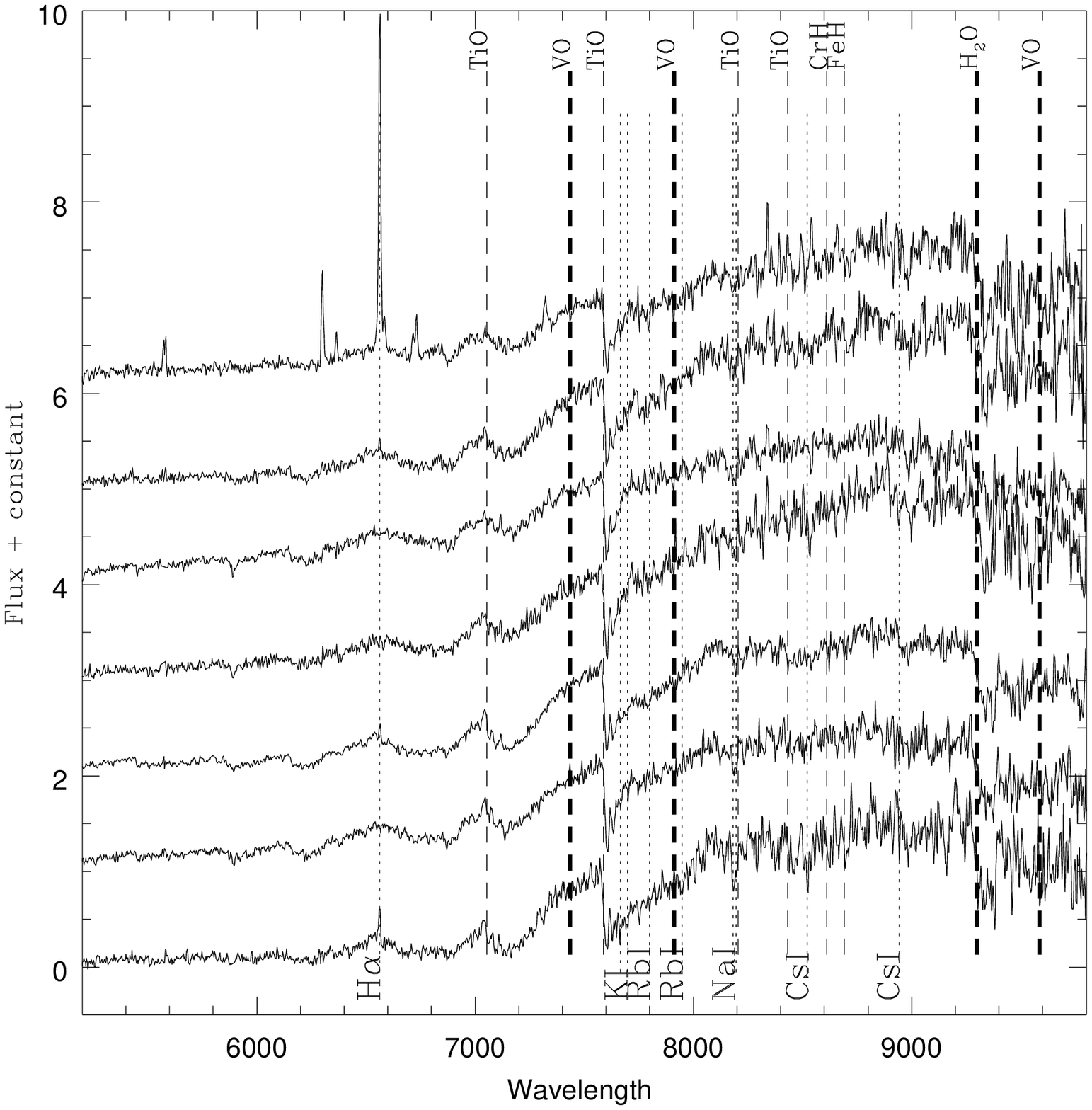}
    \includegraphics[width=10.2cm]{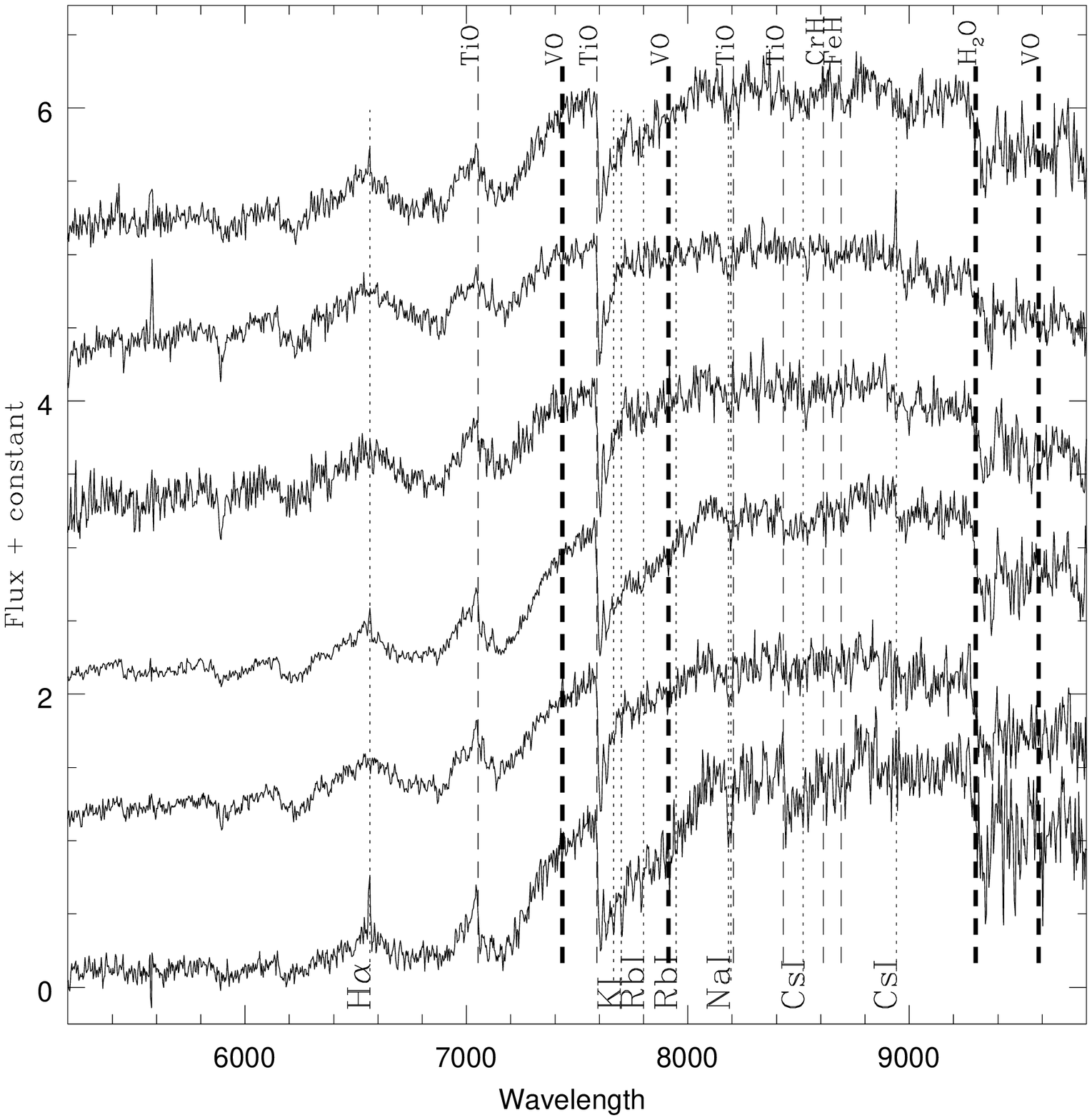}
 \caption{
Spectra taken with the 300 l/mm grating, corresponding to objects
 with moderate reddening.  From bottom to top, [VCE2001] C01, C02, C03, C06, C07, C15 and C41.
{\bf a} Spectra prior dereddening.                    
{\bf b} Spectra after dereddening with our new values --[VCE2001] C41 is not plotted in this case.
}
 \end{figure*}
%______________________________________________________________

\setcounter{figure}{6}
%-----------------------------------------------------------
    \begin{figure*}
    \centering
    \includegraphics[width=16.2cm]{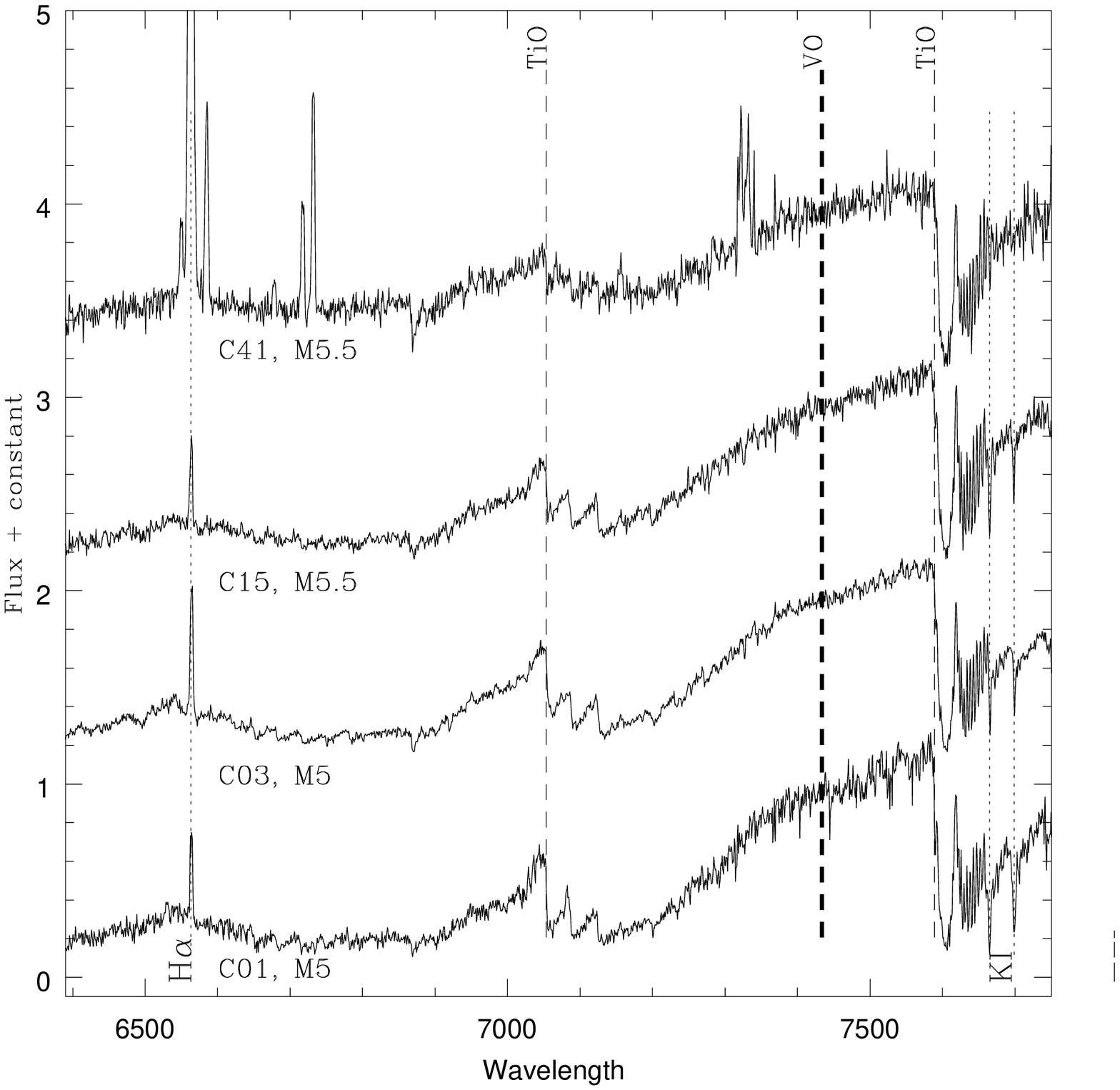}
 \caption{
Spectra taken with the 1200 l/mm grating, corresponding to objects
 with  moderate reddening. Note that the H$\alpha$ emission of [VCE2001] C41 goes
beyond the limits of the figure.
}
 \end{figure*}
%______________________________________________________________

\setcounter{figure}{7}
%-----------------------------------------------------------
    \begin{figure*}
    \centering
    \includegraphics[width=7.8cm]{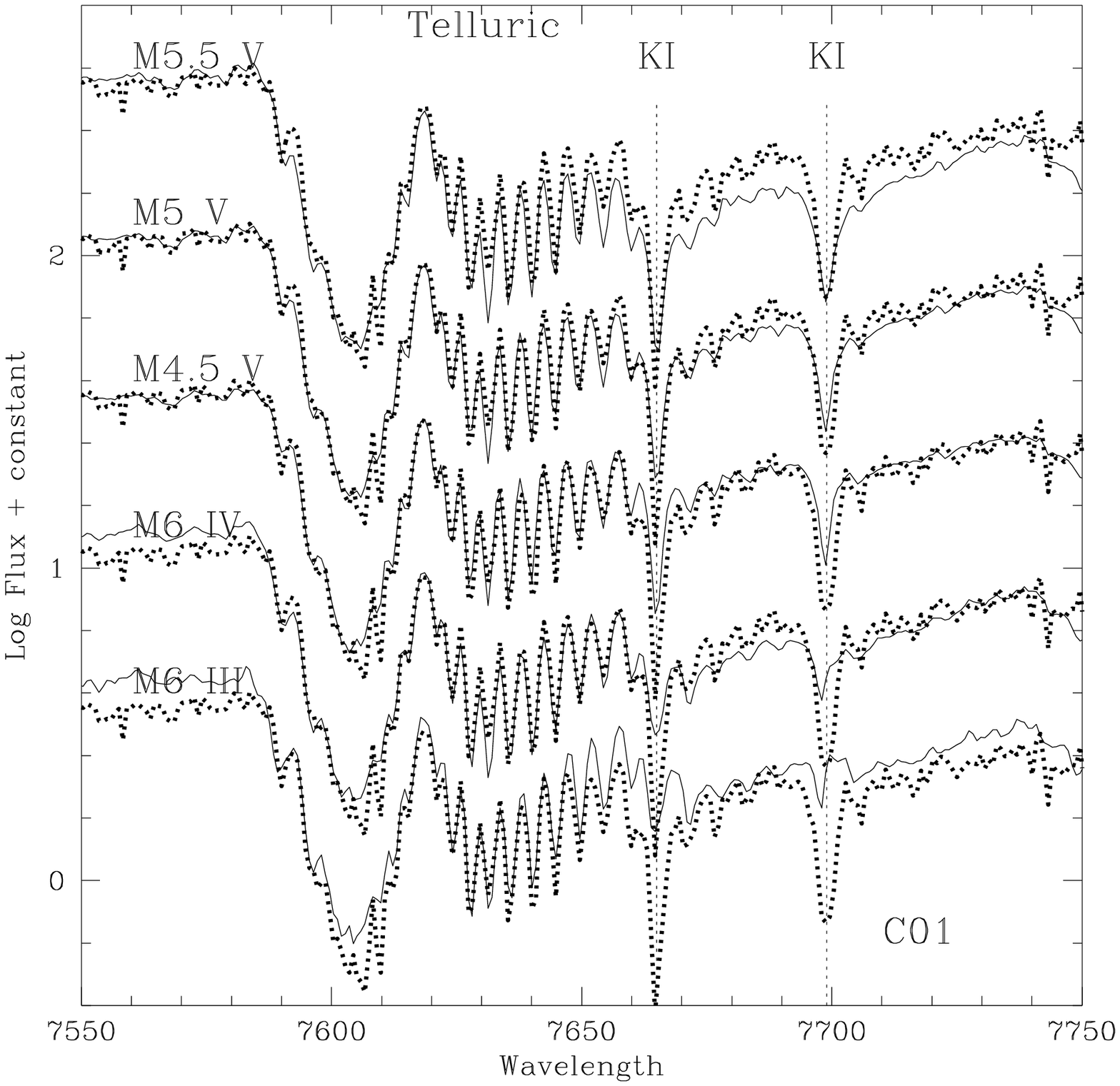}
    \includegraphics[width=7.8cm]{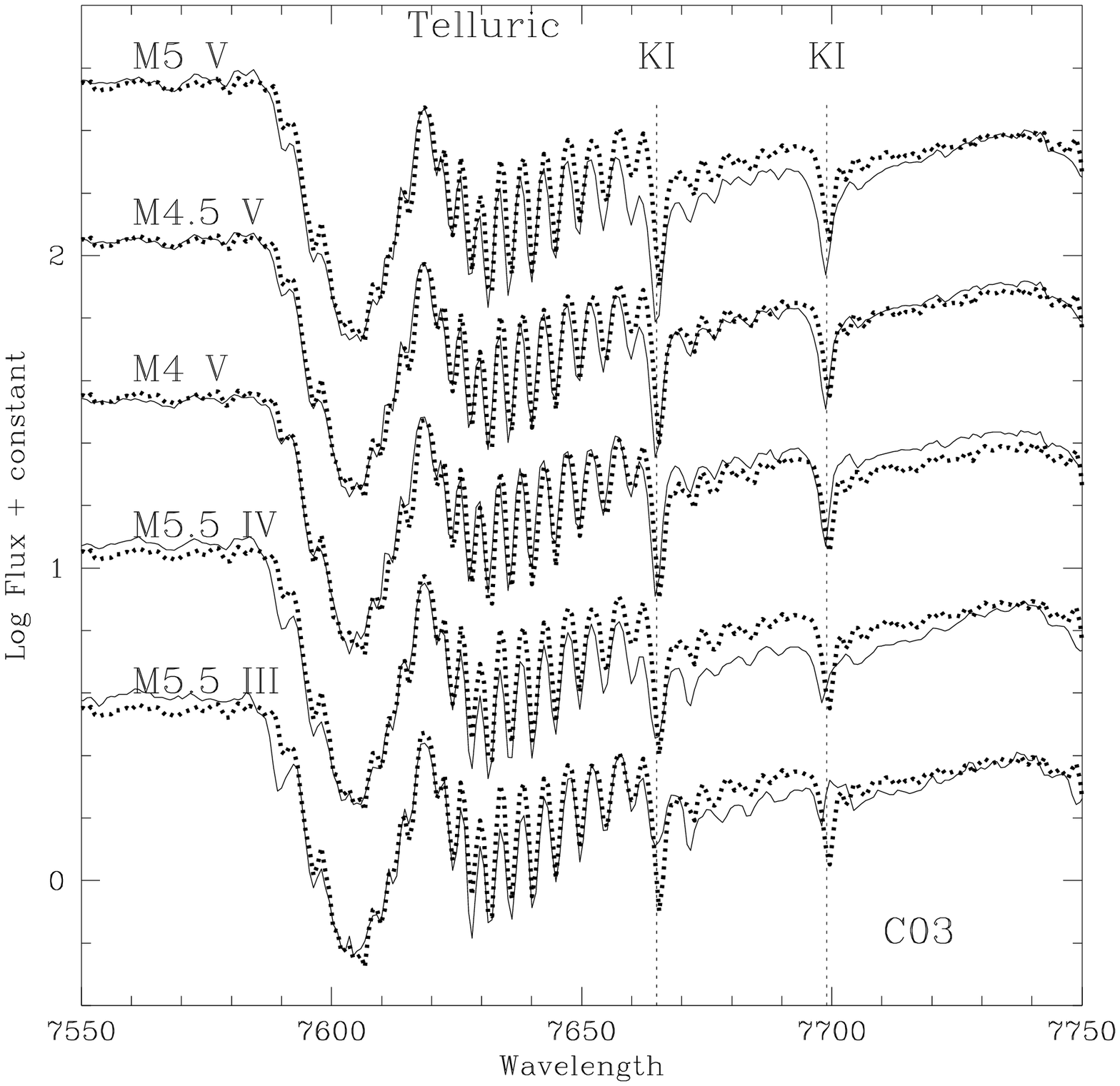}
    \includegraphics[width=7.8cm]{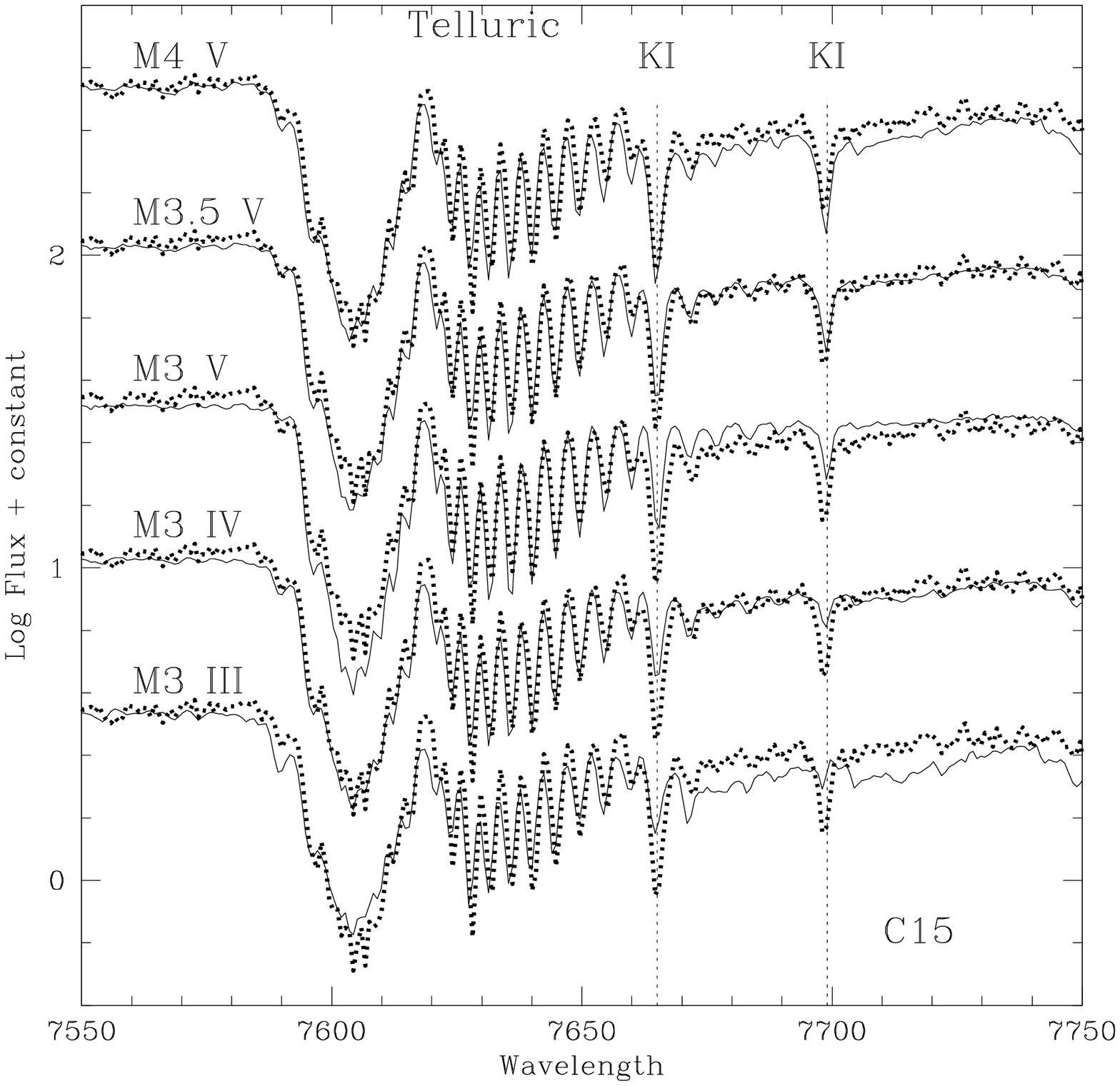}
    \includegraphics[width=7.8cm]{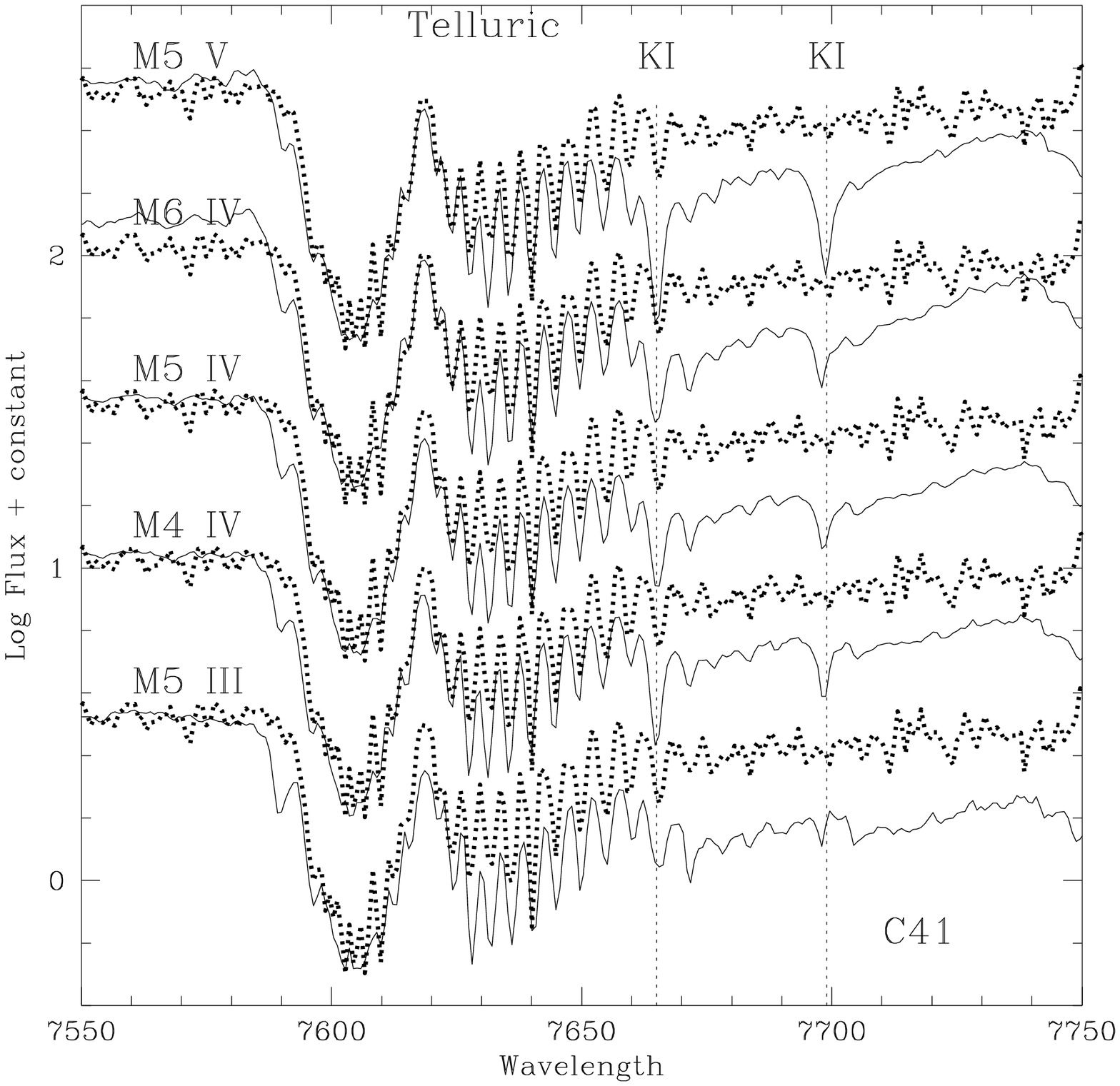}
 \caption{
Spectra around K7700 doublet,
 taken with the 1200 l/mm grating, corresponding to objects
 with none or moderate veiling. Note that some have very strong \ion{K}{1} doublet,
indicating luminosity class V.
{\bf a} [VCE2001] C01,
{\bf b} [VCE2001] C03,
{\bf c} [VCE2001] C15, some veiling might be present in this case.
{\bf d} [VCE2001] C41, the veiling is quite relevant at these wavelenghts.
}
 \end{figure*}
%______________________________________________________________

\setcounter{figure}{8}
%-----------------------------------------------------------
    \begin{figure*}
    \centering
    \includegraphics[width=16.2cm]{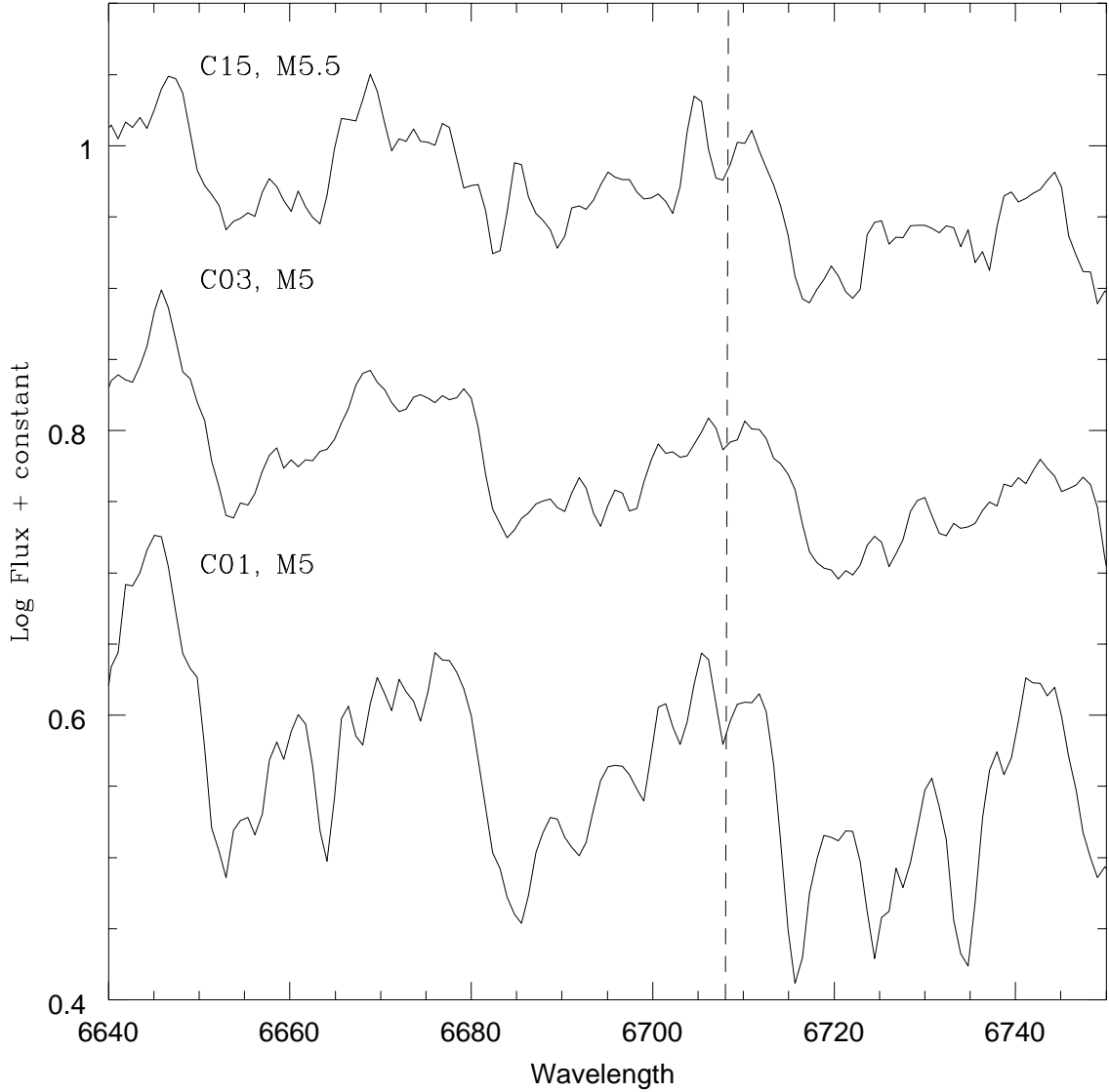}
 \caption{
Area around the \ion{Li}{1} 6707.8 \AA{} doublet. [VCE2001] C01 and C03 lack lithium. 
The depletion is not so clear for [VCE2001] C15, due to the poorer S/N.
This object can suffer some veiling too.
}
 \end{figure*}
%______________________________________________________________

\setcounter{figure}{9}
%-----------------------------------------------------------
    \begin{figure*}
    \centering
    \includegraphics[width=16.2cm]{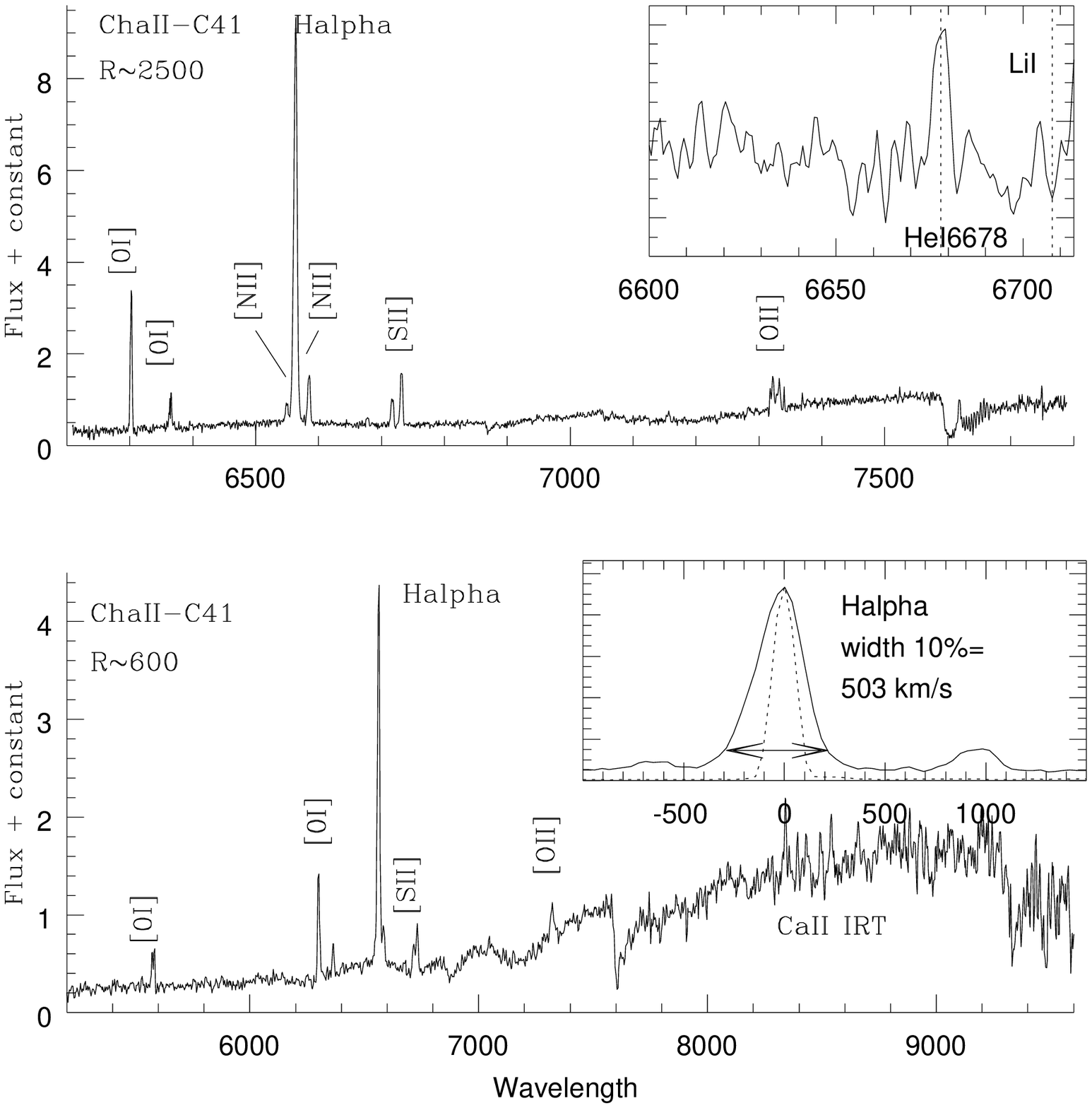}
 \caption{[VCE2001] C41.
{\bf a} Upper panel.- Medium resolution spectrum.
 The zoom in the upper-righ side displays the area around LiI6008 \AA, which 
is detected. 
%
%Note that the doublet \ion{[S]}{2} 6714\&6729 \AA, out of the
%range, is very close.
\ion{He}{1} 6678 \AA{ } also appears in emission. 
{\bf b} Lower panel.- Low resolution spectrum. Note the forbidden
 emission lines. The zoom  corresponds to the area around
H$\alpha$ and \ion{[N]}{2} 6548\&6581 \AA{} at medium resolution (the dotted lines
is the instrumental profile, from the comparison spectrum).
}
 \end{figure*}
%______________________________________________________________

\end{document}